\documentclass{aa}

\usepackage{natbib}
\usepackage{txfonts}
\usepackage[dvips]{graphicx}
\usepackage{color}

\bibpunct{(}{)}{;}{a}{}{,}

\begin{document}

\def\kms{km\,s$^{-1}$}
\def\vsini{v$\sin{i}$}
\def\Vbroad{$V_{\rm broad}$}
\def\teff{$T_{\rm eff}$}
\def\logg{$\log{g}$}
\def\loggf{$\log{gf}$}
\def\C2{${\rm C}_2$}
\def\bv{$B-V$}
\def\mc{\multicolumn}
\def\Msun{${\rm M}_{\sun}$}
\def\Ulsr{$U_{\rm LSR}$}
\def\Vlsr{$V_{\rm LSR}$}
\def\Wlsr{$W_{\rm LSR}$}
\def\Tc{$T_{\rm C}$}

\titlerunning{Abundances in dwarfs, subgiants and giants with and without planets}
\authorrunning{R. da Silva, et al.}

\title{Homogeneous abundance analysis of dwarf, subgiant and giant FGK stars with and without giant planets
\thanks{Based on public data from the ELODIE archive (Moultaka et al. 2004, online access: http://atlas.obs-hp.fr/elodie/)}$^,$
\thanks{Tables~4-8 are only available in electronic form at the CDS via anonymous ftp to cdsarc.u-strasbg.fr (130.79.128.5) or via http://cdsweb.u-strasbg.fr/cgi-bin/qcat?J/A+A/}}

\author{Ronaldo da Silva\inst{1,2} \and Andr\'e de C. Milone\inst{1}
\and Helio J. Rocha-Pinto\inst{3}}

\offprints{R. da Silva,\\
\email{ronaldo.dasilva@roma2.infn.it}}

\institute{
Divis\~ao de Astrofis\'ica, Instituto Nacional de Pesquisas Espaciais, S\~ao Jos\'e dos Campos, Brazil
\and
Dipartimento di Fisica, Universit\`a di Roma Tor Vergata, Rome, Italy
\and
Observat\'orio do Valongo, Universidade Federal do Rio de Janeiro, Rio de Janeiro, Brazil
}

\date{Received / accepted}

%
%
\abstract{}
{We have analyzed high-resolution and high signal-to-noise ratio optical spectra of nearby FGK stars with and without detected giant planets in order to homogeneously measure their photospheric parameters, mass, age, and the abundances of volatile (C, N, and O) and refractory (Na, Mg, Si, Ca, Ti, V, Mn, Fe, Ni, Cu, and Ba) elements. Our sample contains 309 stars from the solar neighborhood (up to the distance of 100~pc), out of which 140 are dwarfs, 29 are subgiants, and 140 are giants.}
{The photospheric parameters are derived from the equivalent widths of \ion{Fe}{i} and \ion{Fe}{ii} lines. Masses and ages come from the interpolation in evolutionary tracks and isochrones on the HR diagram. The abundance determination is based on the equivalent widths of selected atomic lines of the refractory elements and on the spectral synthesis of ${\rm C}_2$, CN, \ion{C}{i}, \ion{O}{i}, and \ion{Na}{i} features. We apply a set of statistical methods to analyze the abundances derived for the three subsamples.}
{Our results show that:
$i)$ giant stars systematically exhibit underabundance in [C/Fe] and overabundance in [N/Fe] and [Na/Fe] in comparison with dwarfs, a result that is normally attributed to evolution-induced mixing processes in the envelope of evolved stars;
$ii)$ for solar analogs only, the abundance trends with the condensation temperature of the elements are correlated with age and anticorrelated with the surface gravity, which is in agreement with recent studies;
$iii)$ as in the case of [Fe/H], dwarf stars with giant planets are systematically enriched in [X/H] for all the analyzed elements, except for O and Ba (the former due to limitations of statistics), confirming previous findings in the literature that not only iron has an important relation with the planetary formation; and
$iv)$ giant planet hosts are also significantly overabundant for the same metallicity when the elements from Mg to Cu are combined together.}
{}

\keywords{stars: fundamental parameters - stars: abundances - planetary systems - methods: data analysis - techniques: spectroscopic}

\maketitle

%
%
\section{Introduction}
\label{intro}

The fact that stars hosting a giant planet are, on average, more abundant in iron than stars in the solar neighborhood for which no planet has been detected \citep{Santosetal2001,Santosetal2004,FischerValenti2005,Gonzalez2006} is well accepted. Actually, it has been shown that this behavior is not exclusive to iron but it is shared by several other metals \citep{Bondetal2006,Gillietal2006,Nevesetal2009,Adibekyanetal2012b}. Other studies also suggested that this kind of anomaly may not only involve the metal content of heavy elements but also some light elements such as carbon, nitrogen, and oxygen \citep{Ecuvillonetal2004,Ecuvillonetal2006,PetiguraMarcy2011}.

Some authors have also been searching for differences in the abundance trends with metallicity for several elements in stars with and without planets. In other words, these works investigated whether the planetary formation in some way affects the observed abundances by comparing stars in the same metallicity range. \citet{Robinsonetal2006} reported that the [Si/Fe] and [Ni/Fe] abundance ratios in planet-host stars are systematically enhanced over their comparison sample of stars without planets of the same metallicity. \citet{DelgadoMenaetal2010} derived the opposite behavior for Mg, and they found no differences regarding [C/Fe], [O/Fe] and [Si/Fe]. In any case, they stated that their result for Mg disappears when only solar analogs are considered. \citet{Nevesetal2009} also reported no significant differences in [X/Fe] for 12 elements in stars with and without planets. \citet{GonzalezLaws2007} derived lower [Al/Fe] and [Si/Fe] but higher [Ti/Fe] in stars with planets in the range of metal-rich stars. \citet{Brugamyeretal2011}, on the other hand, confirmed the [Si/Fe] enhancements in planet hosts, in spite of no [O/Fe] enhancements. Overabundances of several elements were also reported by \citet{Kangetal2011} and \citet{Adibekyanetal2012a}, but only in the range of metal-poor stars, except for Mn. For this element, \citet{Kangetal2011} found an enhancement in the whole range of metallicity, a result that is not supported by \citet{Adibekyanetal2012a}. The situation is, therefore, vast and varied, and also controversial for some elements.

In a previous paper \citep[][from now on referred to as Paper~1]{daSilvaetal2011} we determined photospheric parameters and carbon abundances for a sample of 172 dwarf, subgiant, and giant stars (out of which 18 with planets) using spectra available in the ELODIE database \citep{Moultakaetal2004}. Our previous results did not point out, for instance, any significant difference between the carbon abundances of stars with and without planets, for the three substellar groups. However, the analyzed sample was quite small to provide conclusive results, and stimulated the inclusion of additional stars and other chemical elements.

\begin{figure*}[t!]
\centering
\begin{minipage}[t]{0.33\textwidth}
\centering
\resizebox{\hsize}{!}{\includegraphics{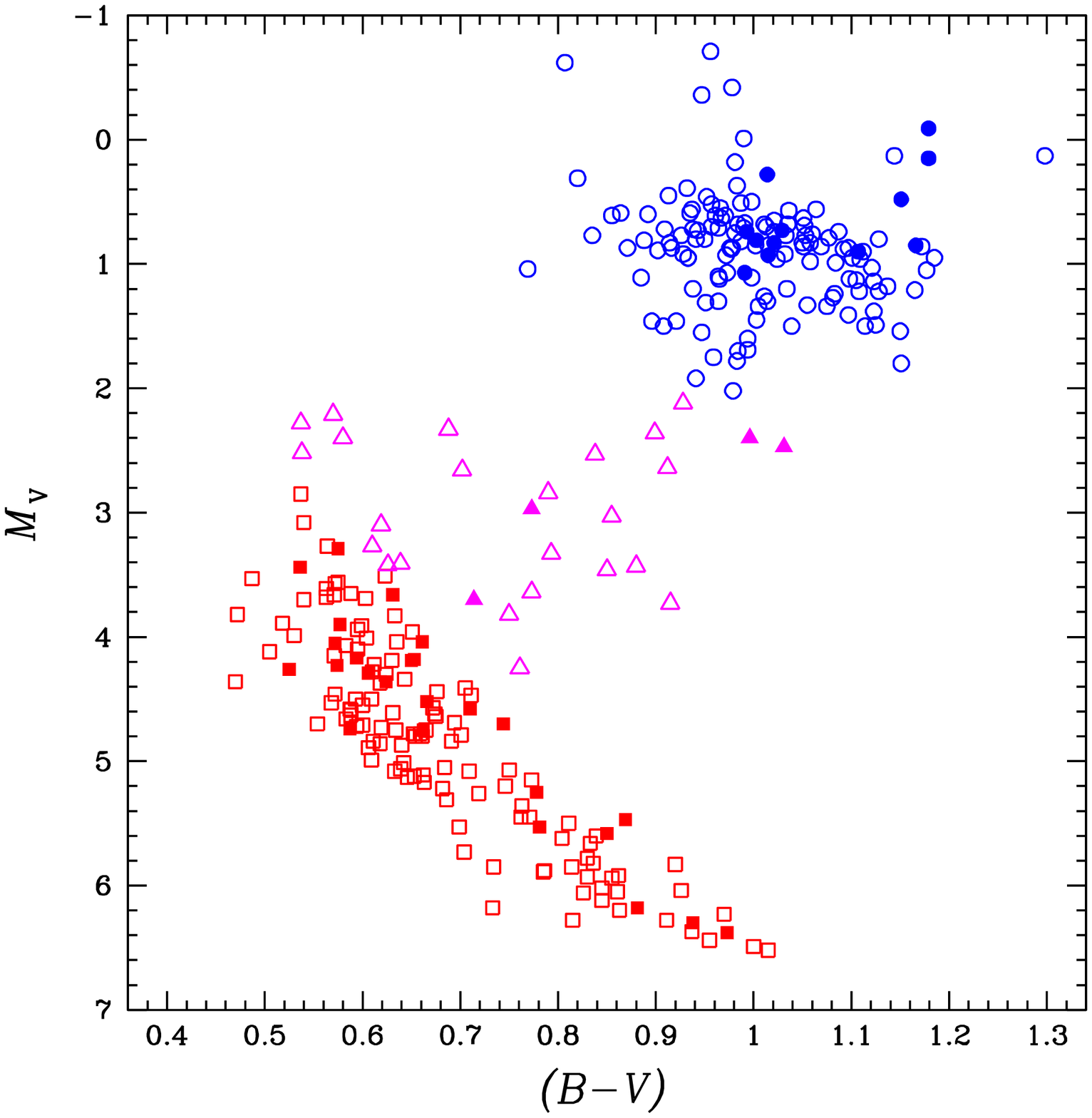}}
\end{minipage}
\begin{minipage}[t]{0.33\textwidth}
\centering
\resizebox{\hsize}{!}{\includegraphics{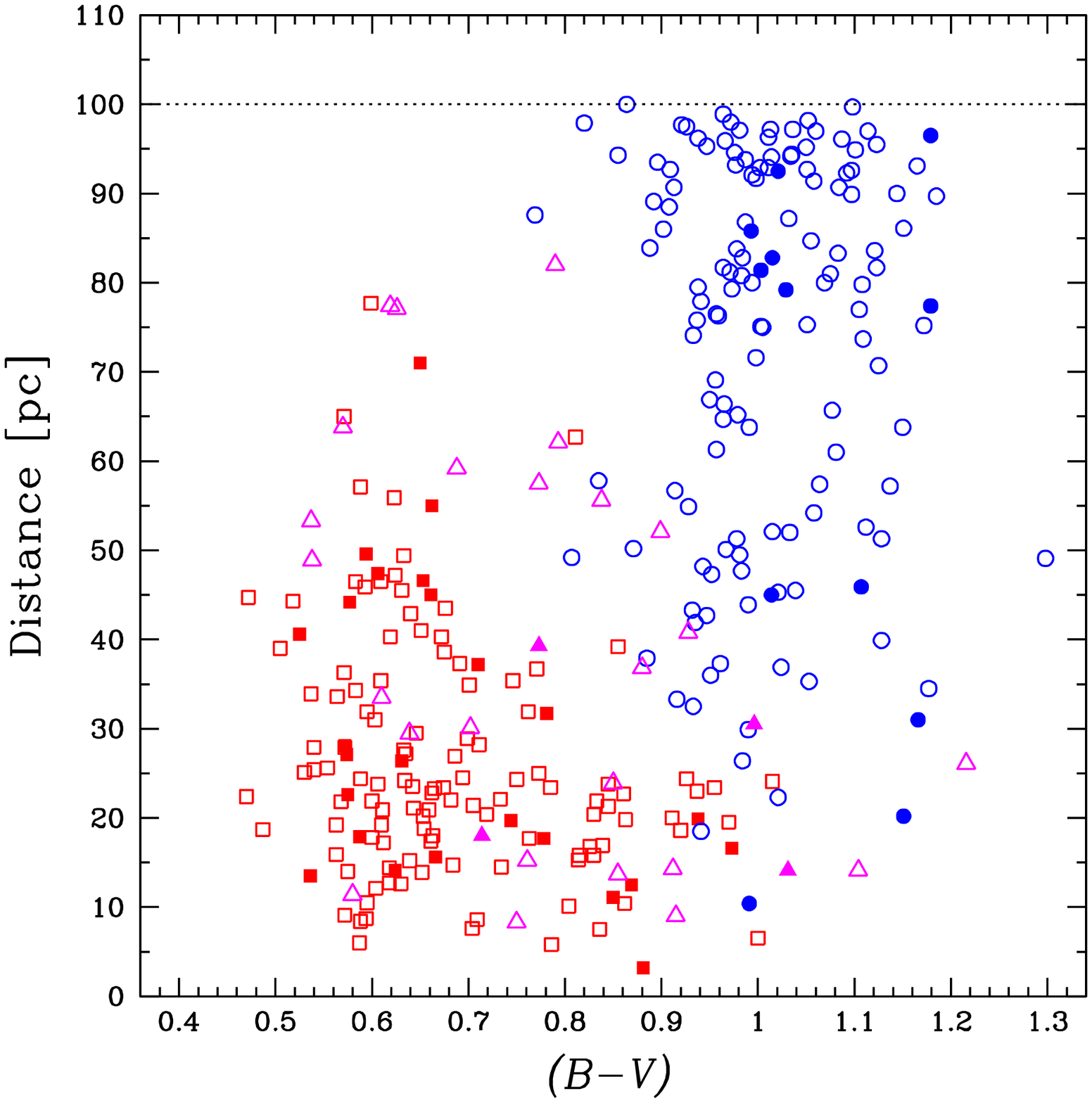}}
\end{minipage}
\begin{minipage}[t]{0.33\textwidth}
\centering
\resizebox{\hsize}{!}{\includegraphics{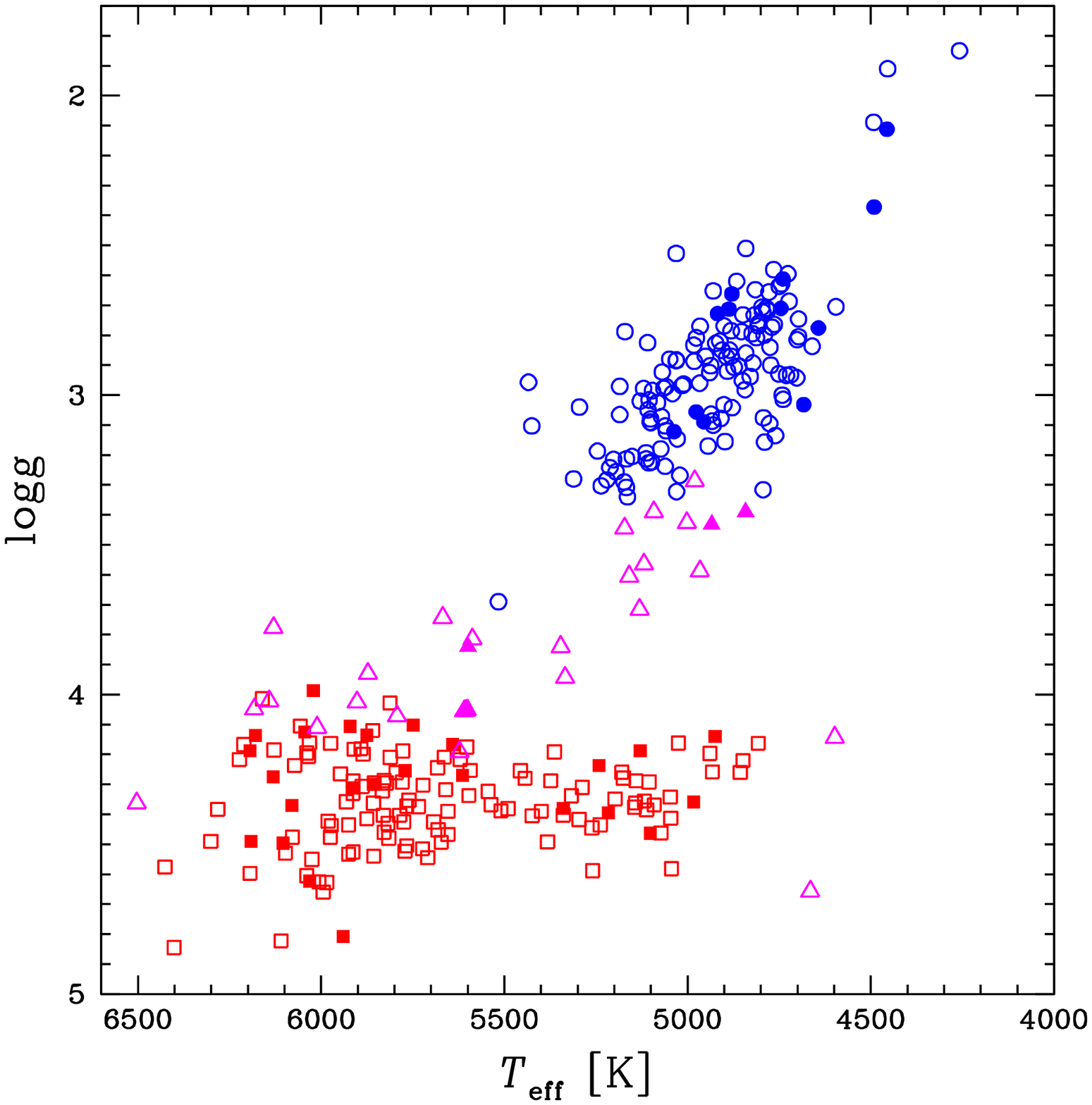}}
\end{minipage}
\caption{HR (left panel) and color-distance (middle panel) diagrams for our sample of 140 dwarfs ({\color{red} $\square$}), 29 subgiants ({\color{magenta} $\triangle$}), and 140 giants ({\color{blue} \Large $\circ$}). The derived stellar parameters are also shown in a \logg\ vs. \teff\ diagram (right panel) for comparison. Filled symbols represent stars with detected planets. The distance limit of 100~pc is represented by the dotted line.}
\label{hr_diag}
\end{figure*}

In the current work we have 309 FGK stars (out of which 47 with detected planets) homogeneously analyzed in order to measure their photospheric parameters, masses, ages, and the abundances of volatile (C, N, and O) and refractory (Na, Mg, Si, Ca, Ti, V, Mn, Fe, Ni, Cu, and Ba) elements. The sample includes dwarf, subgiant, and giant stars, mostly belonging to the Galactic thin disc. We have applied a set of statistical methods to the three subsamples, exploring their multivariate properties and, in particular, uncovering significant differences between dwarf stars with and without a known giant planet. In addition, we have searched for relations involving the stellar parameters, the abundances, and the condensation temperature of the elements. This approach has been very often used in order to check whether any difference in the amount of refractory elements with respect to volatile ones has an influence on the formation of planetary systems \citep[see, e.g.,][]{Smithetal2001,Melendezetal2009,Ramirezetal2009,GonzalezHernandezetal2010,GonzalezHernandezetal2013a,GonzalezHernandezetal2013b}.

In Sect.~\ref{obs} we describe the observations and the procedure of data reduction. In Sect.~\ref{spec_an} we outline the methods used to derive the photospheric parameters and the chemical abundances. The determination of the evolutionary and the kinematic parameters is presented in Sect.~\ref{evol_kinem}. The results are presented and discussed in Sect.~\ref{res} and in Sect.~\ref{stat_an}, in which we describe the statistical analysis. Finally, we present our final remarks and conclusions in Sect.~\ref{concl}.

%
%
\section{Observation data and reduction}
\label{obs}

In this work, as a complement to our previous analysis of 172 stars published in Paper~1, we have enlarged our stellar sample. Now we have 309 FGK stars from the solar neighborhood that include 140 dwarfs (out of which 31 with detected planets), 29 subgiants (4 with planets), and 140 giants (12 with planets).

The stars were observed with the ELODIE high-resolution spectrograph \citep{Baranneetal1996} of the Haute Provence Observatory (France), and the data are publicly available in the online ELODIE database. The ELODIE spectrograph provides spectra with resolution $R\sim42\,000$ in the wavelength range 3895$-$6815~\AA. The criteria of selection are similar, but not equal, to those applied before since we have now considered a smaller inferior limit for the signal-to-noise ratio (S/N). Following this new criteria we selected:

\begin{itemize}

\item[\it i)] stars for which the averaged spectra have S/N $\geq$ 150; only individual spectra with a S/N $\geq$ 20 and with an image type classified as {\it object fibre only} (OBJO) were used;

\item[\it ii)] stars within a distance $\leq$ 100~pc (parallax $\pi \geq$ 10~mas) and with spectral type between F8 and M1;

\item[\it iii)] stars having no close-in binary companion that could contaminate the observed spectra; we used the information of the angular separation between components ($rho$) available in the Hipparcos catalogue \citep{ESA1997} and we chose only the systems with $rho >$ 10~arcsec; additionally, we searched in literature for any further information that could unveil a contaminant star in the observed field;

\item[\it iv)] stars with (\bv) measured by Hipparcos and with cross-correlation parameters available in the ELODIE database; both (\bv) and the width of the spectral cross-correlation function are required to carry out a first estimate of the projected rotation velocity \vsini\ of each star; and

\item[\it v)] stars for which the determination of the photospheric parameters is reliable (see Sect.~\ref{phot_par}) and that passed the quality control of the spectral synthesis (see Sect.~\ref{abund}).

\end{itemize}

A total of 1485 spectra of 309 stars has been analyzed. The subsamples of dwarfs, subgiants, and giants are plotted in the HR and color-distance diagrams of Fig.~\ref{hr_diag}. As in Paper~1, we chose to classify as subgiants those stars situated 1.5~mag above the lower limit of the main-sequence and having $M_{\rm v} > 2.0$~mag. Note that the distance of dwarfs and subgiants is not limited to 100~pc, but to about 80~pc, which represents a selection effect of the ELODIE observation surveys. For comparison, in the right panel of this figure we plot the \logg\ vs. \teff\ diagram with the stellar parameters derived in the current work.

\begin{figure*}[t!]
\centering
\begin{minipage}[t]{0.33\textwidth}
\centering
\resizebox{\hsize}{!}{\includegraphics{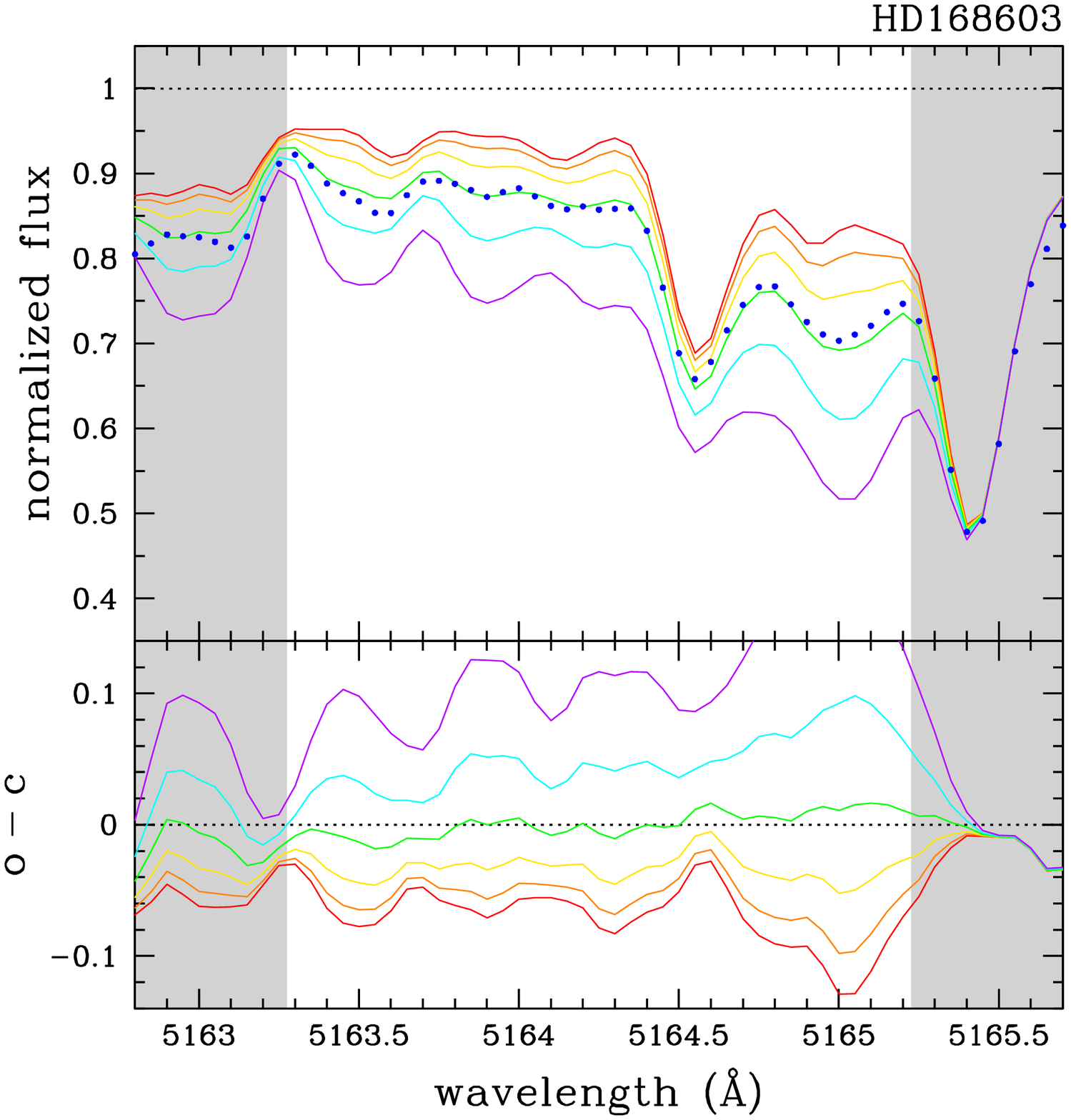}}
\end{minipage}
\begin{minipage}[t]{0.33\textwidth}
\centering
\resizebox{\hsize}{!}{\includegraphics{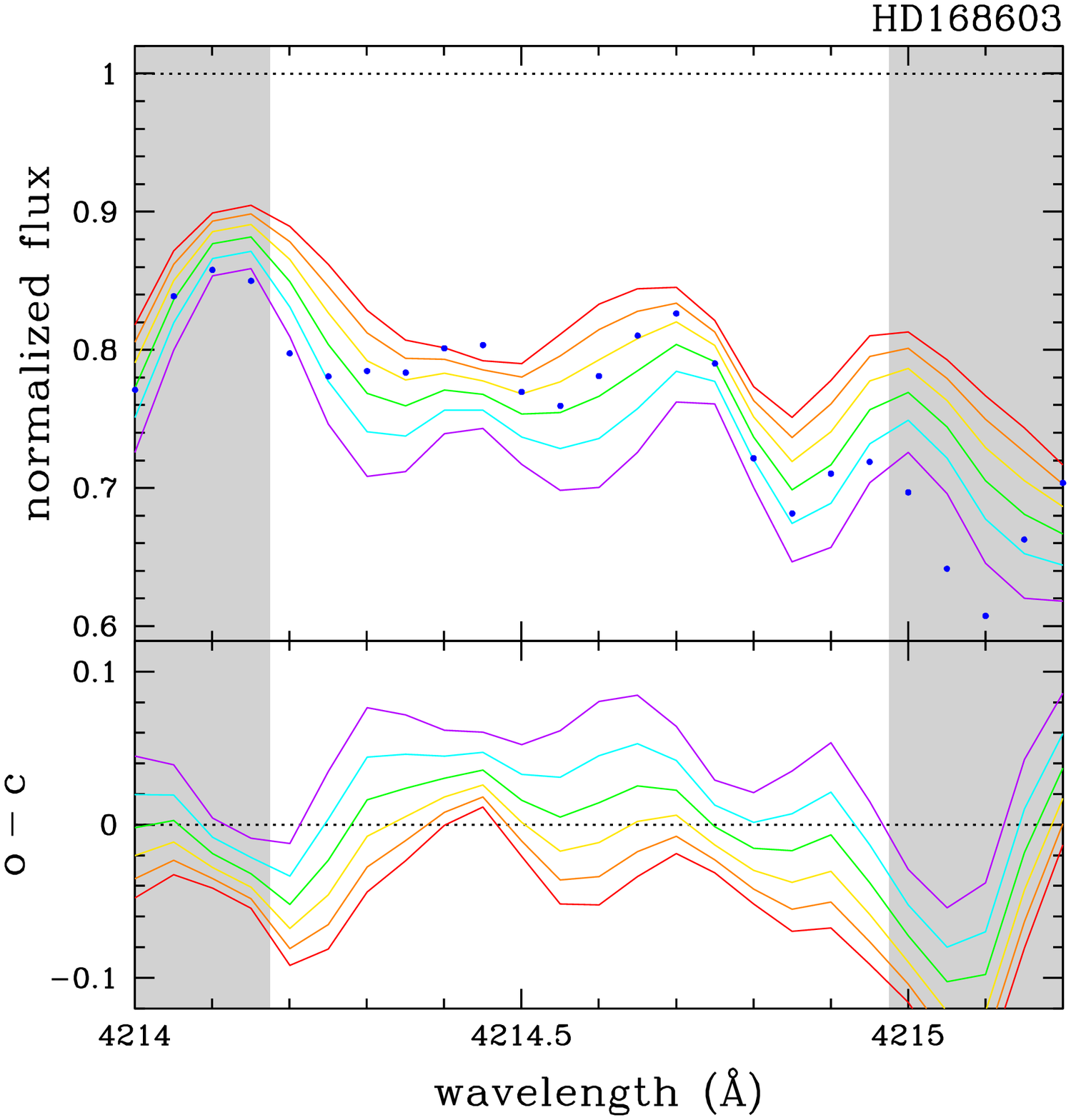}}
\end{minipage}
\begin{minipage}[t]{0.33\textwidth}
\centering
\resizebox{\hsize}{!}{\includegraphics{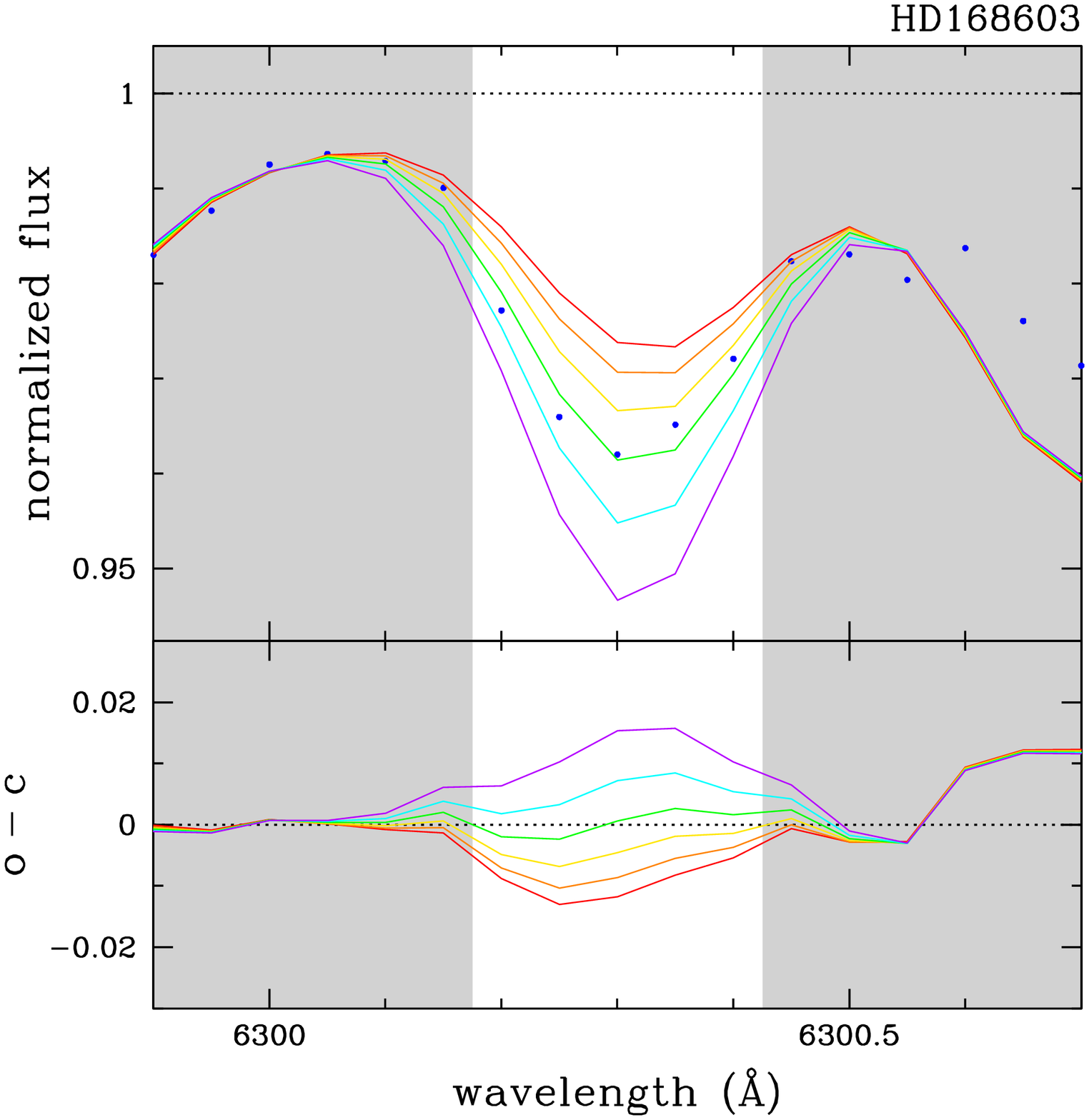}}
\end{minipage} \\
\begin{minipage}[t]{0.33\textwidth}
\centering
\resizebox{\hsize}{!}{\includegraphics{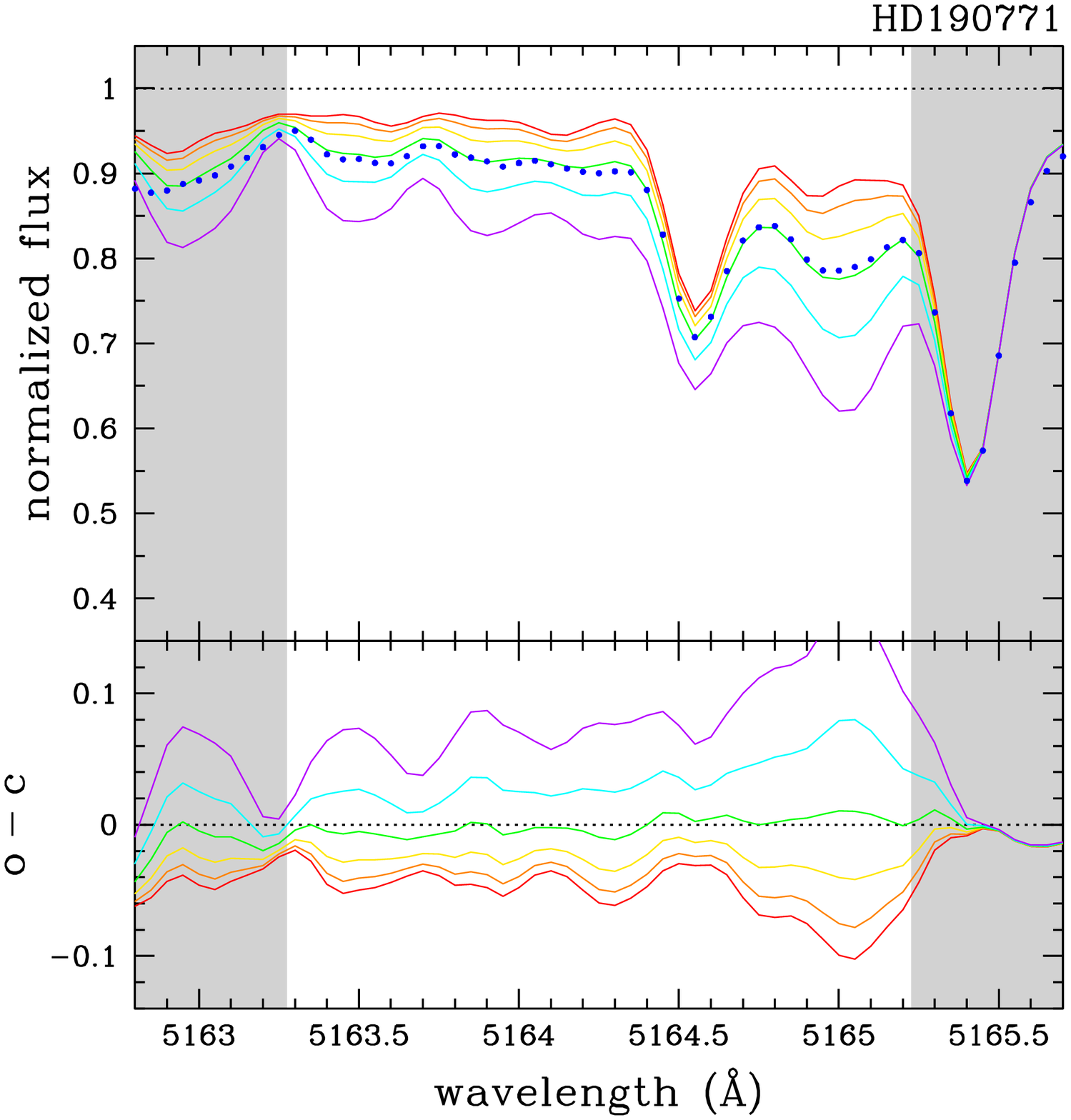}}
\end{minipage}
\begin{minipage}[t]{0.33\textwidth}
\centering
\resizebox{\hsize}{!}{\includegraphics{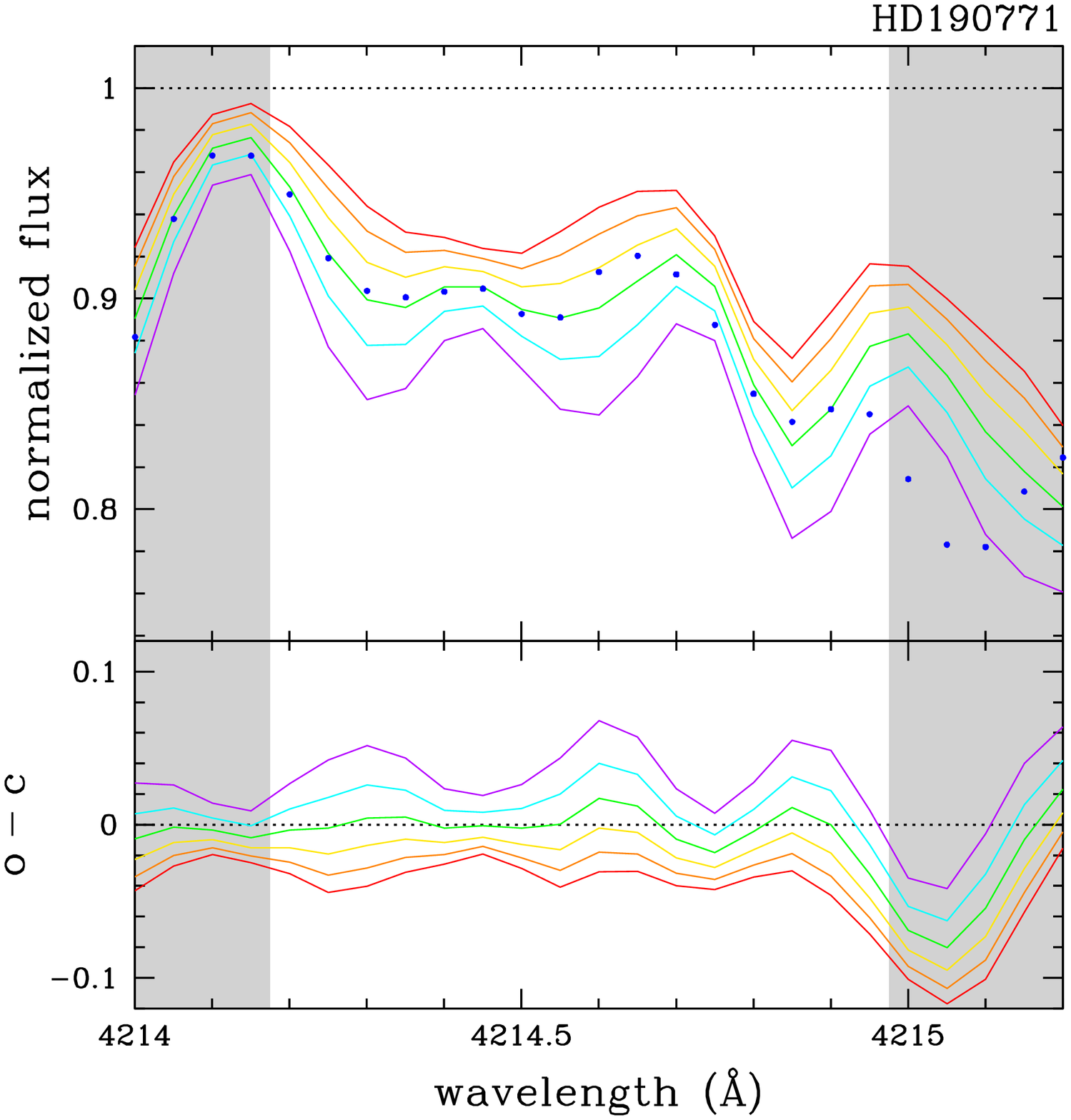}}
\end{minipage}
\begin{minipage}[t]{0.33\textwidth}
\centering
\resizebox{\hsize}{!}{\includegraphics{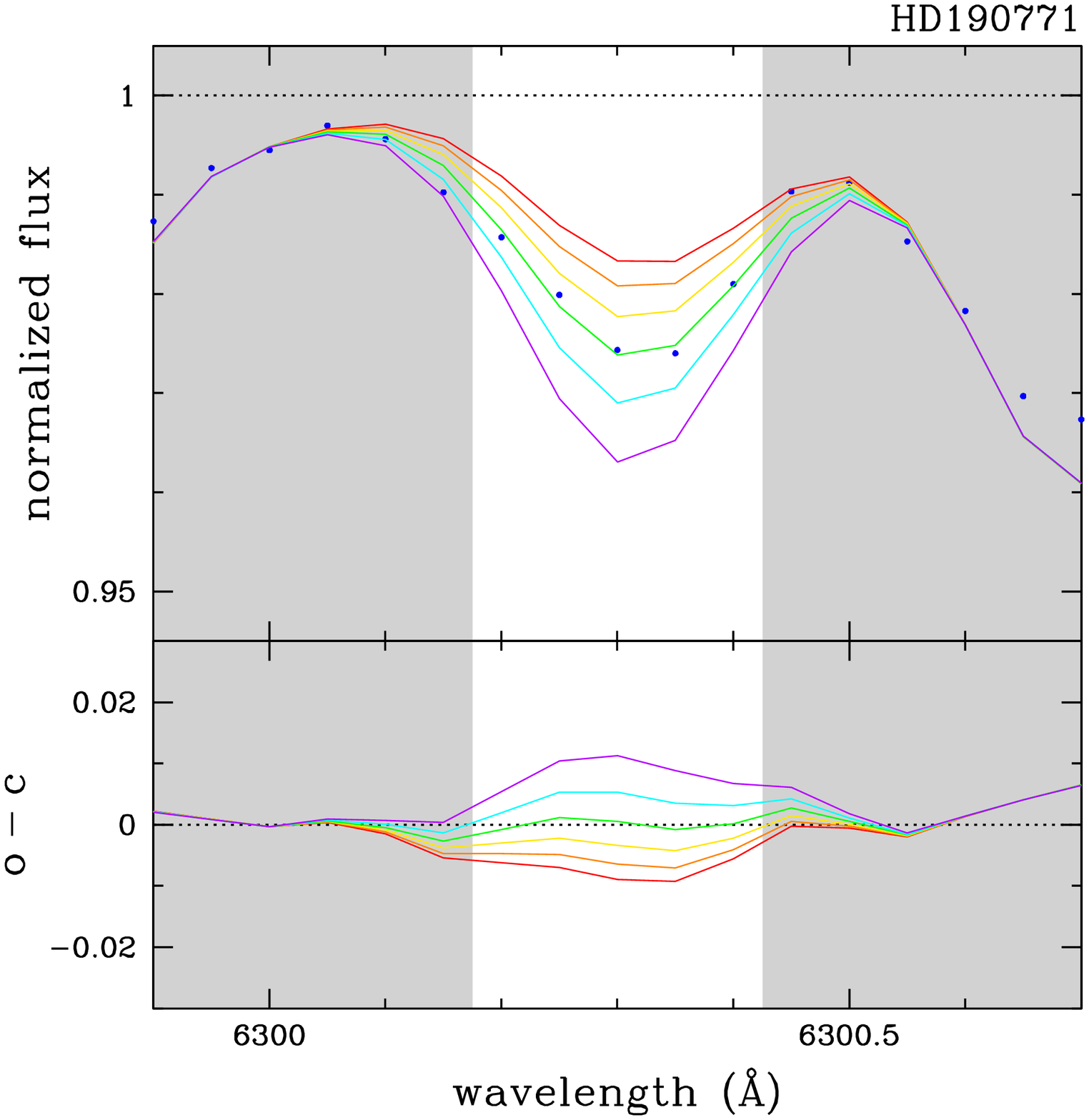}}
\end{minipage}
\caption{Synthesis of three spectral regions: \C2\ ($\lambda$5165, left panels) and CN ($\lambda$4215, middle panels) molecular band heads, and the O atomic line at $\lambda$6300.30 (right panels). Two dwarf stars with different values of spectral S/N are shown: HD\,168603 (S/N = 156, \teff\ = 5399~K, [Fe/H] = 0.11~dex, top panels) and HD\,190771 (S/N = 299, \teff\ = 5817~K, [Fe/H] = 0.14~dex, bottom panels). These are the S/N per pixel computed at $\lambda$5550. Six spectra computed for different abundances (the N abundance is changed in the case of the CN band) and separated by 0.1~dex are plotted. The differences between observed and computed spectra (O$-$C) are also plotted in the bottom of each panel.}
\label{stars_synth}
\end{figure*}

The spectra were reduced using IRAF\,\footnote{{\it Image Reduction and Analysis Facility}, distributed by the National Optical Astronomy Observatories (NOAO), USA.} routines for order identification and extraction, background subtraction, flat-field correction, wavelength calibration, radial-velocity shift correction, and flux normalization (a global normalization first, and then a more careful one performed around regions containing molecular bands and atomic lines used in the spectral synthesis).

%
%
\section{Spectroscopic analysis}
\label{spec_an}

\subsection{Photospheric parameters and equivalent widths}
\label{phot_par}

In this work, as in Paper~1, we performed a differential analysis relative to the Sun. The method requires a model atmosphere for each star, which in turn is computed using the so-called photospheric parameters: effective temperature \teff, metallicity [Fe/H], surface gravity \logg, and microturbulent velocity $\xi$. We computed these parameters through the excitation equilibrium of neutral iron and the ionization equilibrium between \ion{Fe}{i} and \ion{Fe}{ii} lines, using model atmospheres derived by interpolation in the grid of \citet{Kurucz1993}. As before, we used the MOOG LTE radiative code \citep{Sneden2002} and the same routine developed by us to perform an automated calculation of the parameters and their uncertainties, allowing a fast and homogeneous approach.

To compute the solar model atmosphere we adopted \teff\ = 5777~K, \logg\ = 4.44, $\xi$ = 1.0~\kms, and the standard solar abundances of \citet{AndersGrevesse1989}, but with $\log{\epsilon}_{\,{\sun}}(\rm Fe)$ = 7.47 to be consistent with the value adopted in Paper~1. We also assumed the Uns\"old approximation multiplied by 6.3 to account for the van der Waals line damping. This is useful for a better treatment of strong lines close to the spectral windows used in our spectral synthesis analysis.

\begin{figure*}
\centering
\begin{minipage}[b]{\textwidth}
\centering
\resizebox{\hsize}{!}{\includegraphics{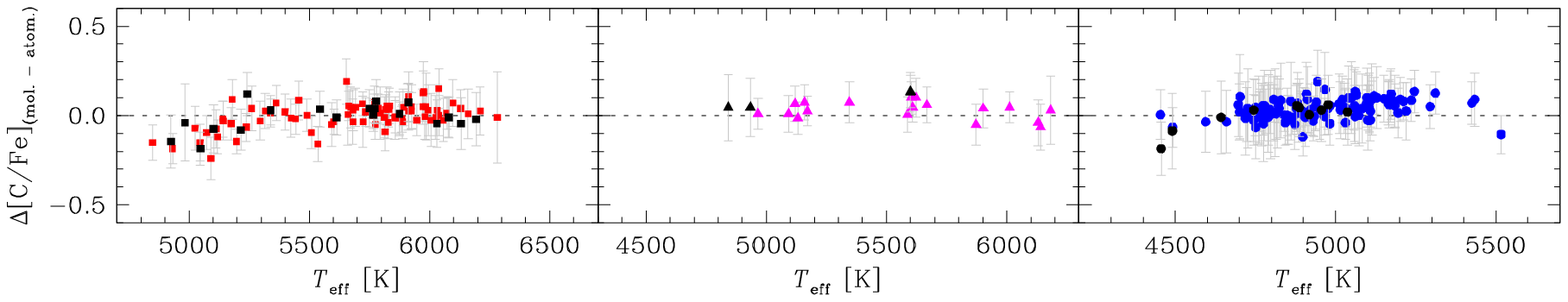}}
\end{minipage} \\
\begin{minipage}[b]{\textwidth}
\centering
\resizebox{\hsize}{!}{\includegraphics{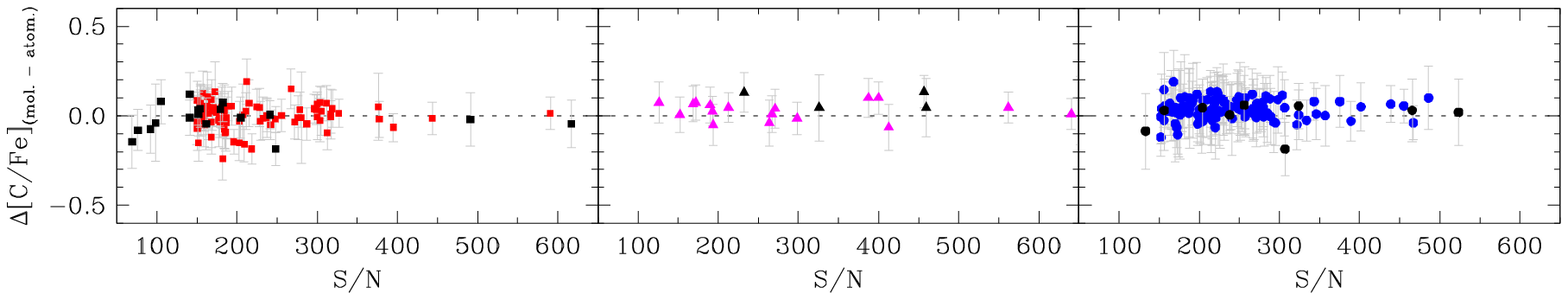}}
\end{minipage}
\caption{Difference between the C abundances derived from the \C2\ molecular bands ($\lambda$5128 and $\lambda$5165) and the C atomic line ($\lambda$5380.34) as a function of the effective temperature (top panels) and of the S/N (bottom panels) for our sample of dwarfs (left panels), subgiants (middle panels), and giants (right panels). Stars with planets are represented by black symbols.}
\label{Dcfe_teff_sn}
\end{figure*}

The atomic line parameters (wavelength, statistical weight of the lower level multiplied by the oscillator strength $gf$, and lower-level excitation potential $\chi$) of 71 \ion{Fe}{i} and 12 \ion{Fe}{ii} lines are listed in Table~\ref{linelist}. The table also lists the parameters of the other elements analyzed by means of their equivalent width ($EW$). Table~\ref{hfs} contains the $gf$ values of the elements with important hyperfine structure (HFS). These parameters were initially taken from the {\it Vienna Atomic Line Database} -- VALD \citep{Kupkaetal2000,Kupkaetal1999,Piskunovetal1995,Ryabchikovaetal1997} in the case of the elements without HFS, and from \citet{Steffen1985} for those with important HFS. The $gf$ values were then revised in order to fit the equivalent widths measured in the spectrum of the Solar Flux Atlas \citep{Kuruczetal1984} degraded to the ELODIE resolution. For the Sun and for all the other stars of our sample, the $EW$s were measured using the {\it Automatic Routine for line Equivalent widths in stellar Spectra} -- ARES \citep{Sousaetal2007}, and they are also listed in Table~\ref{linelist}.

The atomic lines of the elements Mg, V, Mn, and Cu have important HFS that has to be taken into account. In the case of those elements with lines not listed in \citet{Steffen1985}, we adopted the values of neighboring multiplets. An exception is the Mg line at $\lambda$5785.29, for which no HSF information was available. Its $gf$ value was obtained in the same way as those lines without HFS listed in Table~\ref{linelist}. This line is not strong ($EW < 60$ m\AA) and the error induced in the Mg abundance determination is of second order. The HFS of Ba, and also its isotopic splitting, are of some importance but only for the line at $\lambda$6496.91, being negligible for the lines at $\lambda$5853.69 and $\lambda$6141.73 \citep[see][]{Korotinetal2011}.

Table~\ref{spec_pars} shows an excerpt of the derive photospheric parameters for the 140 dwarf stars in our sample, together with their uncertainties. The complete tables, also for subdwarfs and giants, are available in electronic form at the CDS.

\subsection{Abundance determination}
\label{abund}

The abundances of the refractory elements (Na, Mg, Si, Ca, Ti, V, Mn, Ni, Cu, and Ba), with respect to Fe, were determined based on the equivalent widths of a selection of spectral lines chosen as being free from blends by inspection of the Solar Flux Atlas and the solar line identifications catalogue of \citet{Mooreetal1966}.

We used the {\it abfind} (in the case of elements without important HFS), and the {\it blends} (in the case of elements with HFS) drivers of the MOOG code to derive the abundances. These drivers require:
$(i)$ the model atmospheres, which were previously computed using the photospheric parameters derived for each star; and
$(ii)$ a list of the selected lines, containing the respective wavelength (or the hyperfine structure), the atomic number, the lower-level excitation potential, the revised values of $gf$, and the measured equivalent widths. Again, using another routine for an automated calculation developed by us, we derived the abundances for each element in turn, excluding when needed (by means of a sigma-clipping procedure) spectral lines that provided a bad abundance determination.

\begin{figure*}
\centering
\begin{minipage}[b]{0.33\textwidth}
(a)\hspace{2.5cm}\,\\[-4cm]
\centering
\resizebox{\hsize}{!}{\includegraphics[angle=-90]{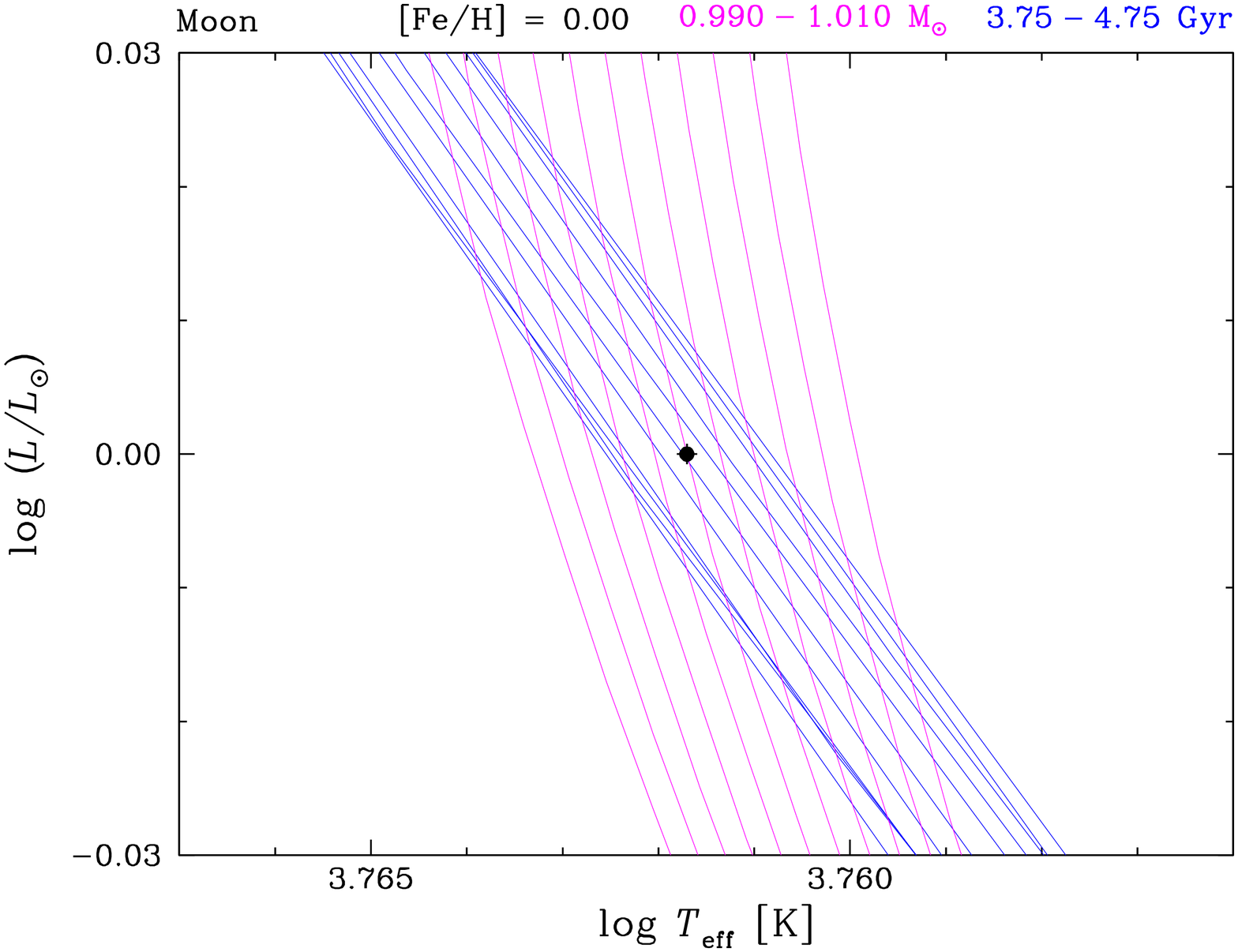}}
\end{minipage}
\begin{minipage}[b]{0.33\textwidth}
(b)\hspace{2.5cm}\,\\[-4cm]
\centering
\resizebox{\hsize}{!}{\includegraphics[angle=-90]{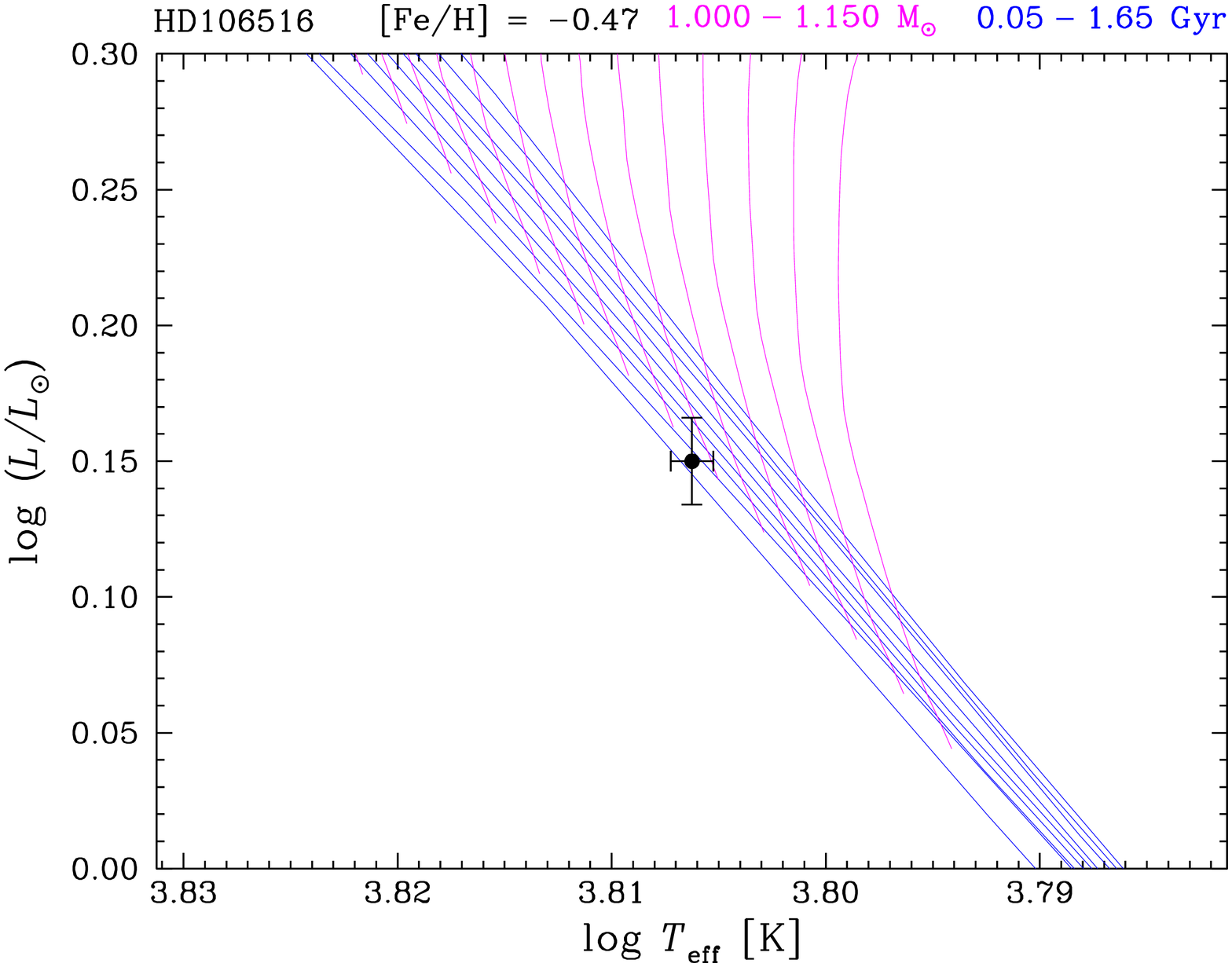}}
\end{minipage}
\begin{minipage}[b]{0.33\textwidth}
(c)\hspace{2.5cm}\,\\[-4cm]
\centering
\resizebox{\hsize}{!}{\includegraphics[angle=-90]{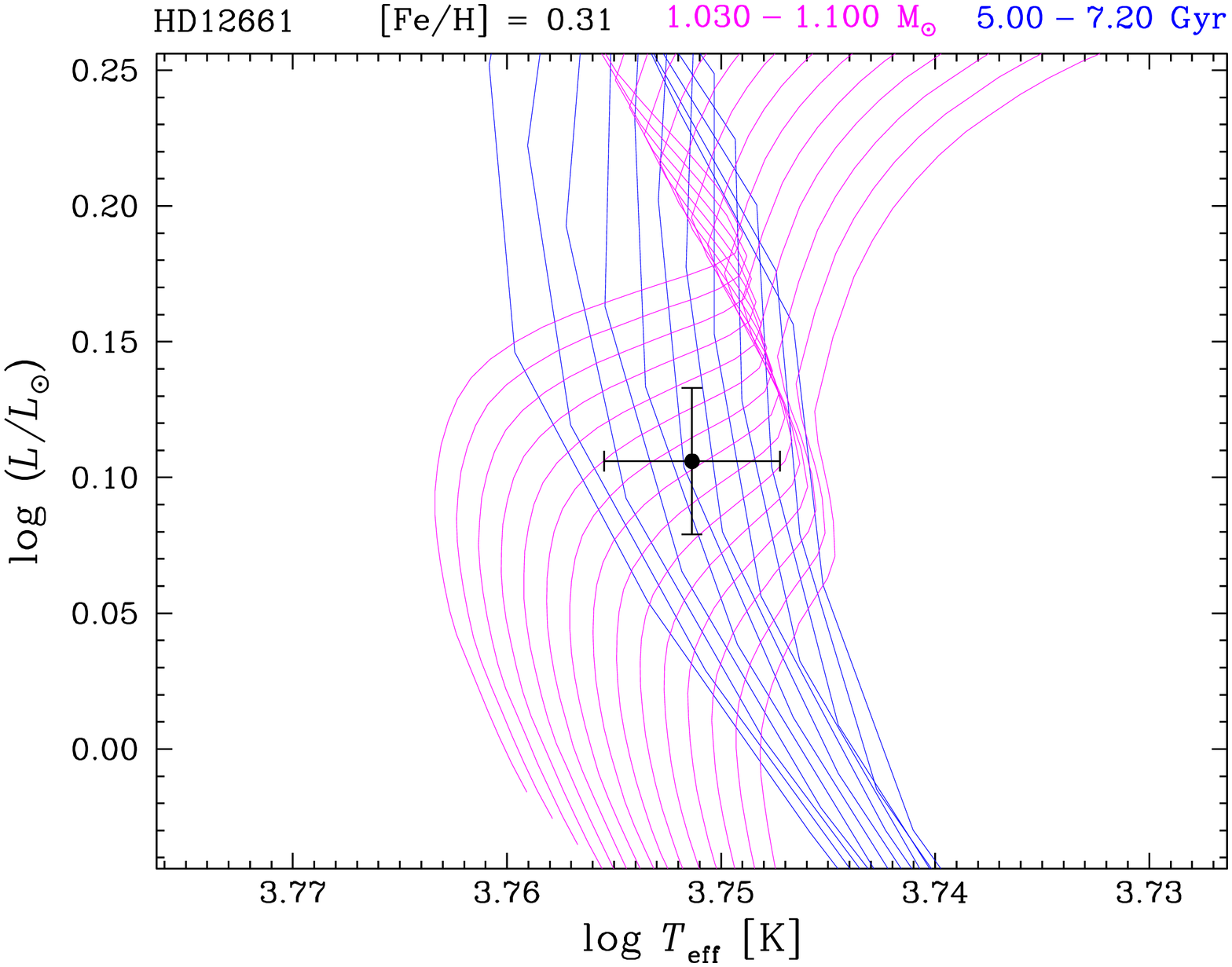}}
\end{minipage} \\
\begin{minipage}[b]{0.33\textwidth}
(d)\hspace{2.5cm}\,\\[-4cm]
\centering
\resizebox{\hsize}{!}{\includegraphics[angle=-90]{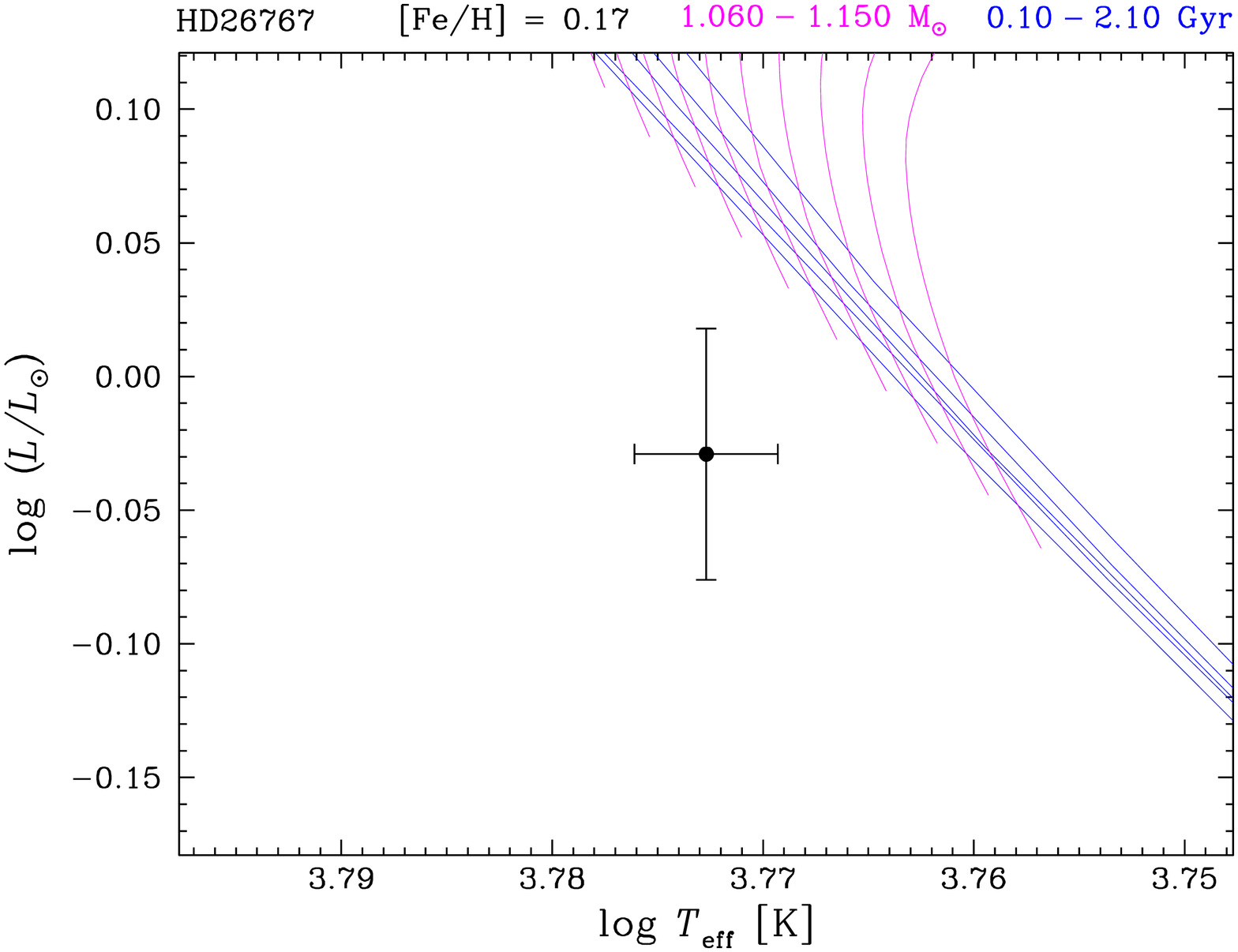}}
\end{minipage}
\begin{minipage}[b]{0.33\textwidth}
(e)\hspace{2.5cm}\,\\[-4cm]
\centering
\resizebox{\hsize}{!}{\includegraphics[angle=-90]{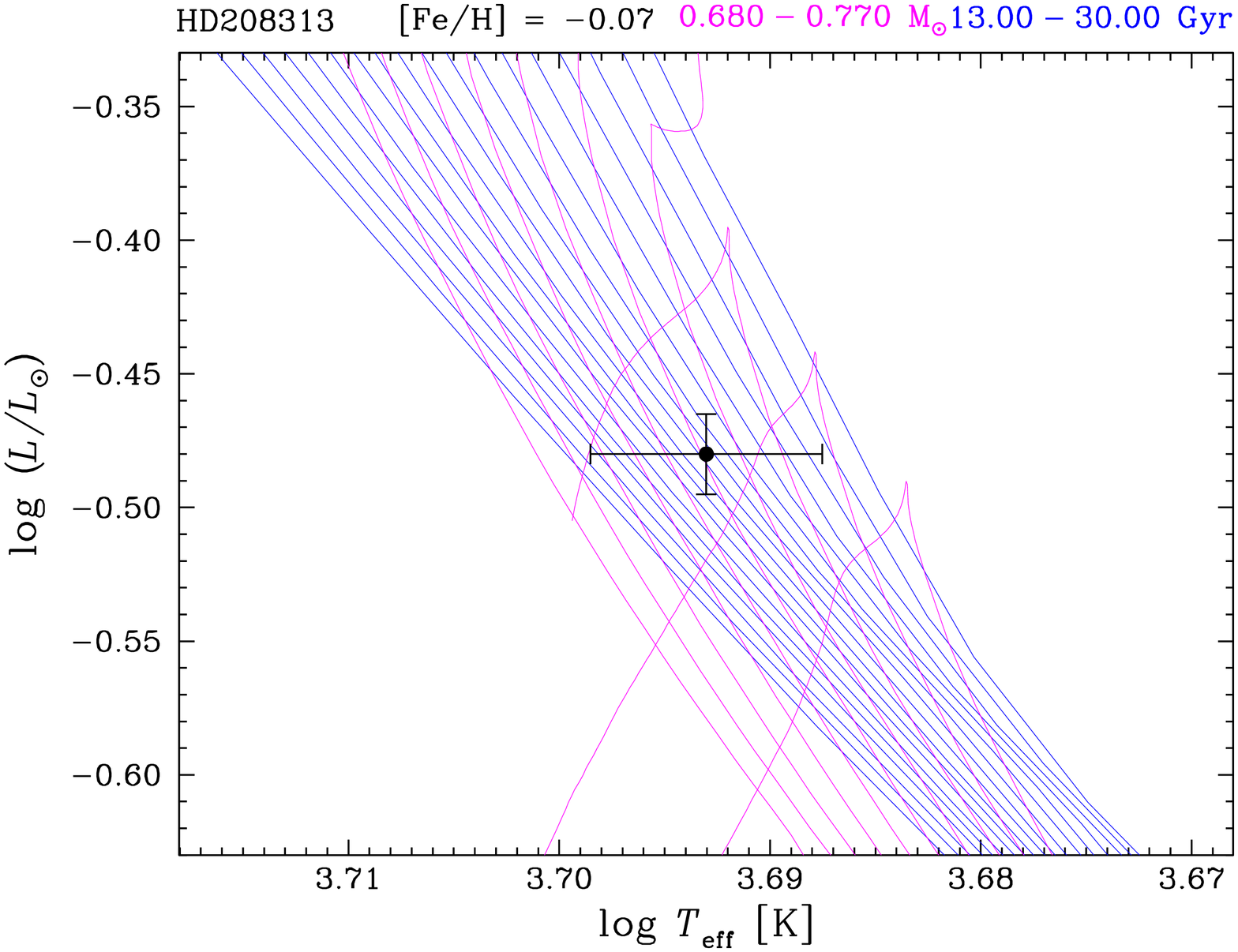}}
\end{minipage}
\begin{minipage}[b]{0.33\textwidth}
(f)\hspace{-9.5cm}\,\\[-4cm]
\centering
\resizebox{\hsize}{!}{\includegraphics[angle=-90]{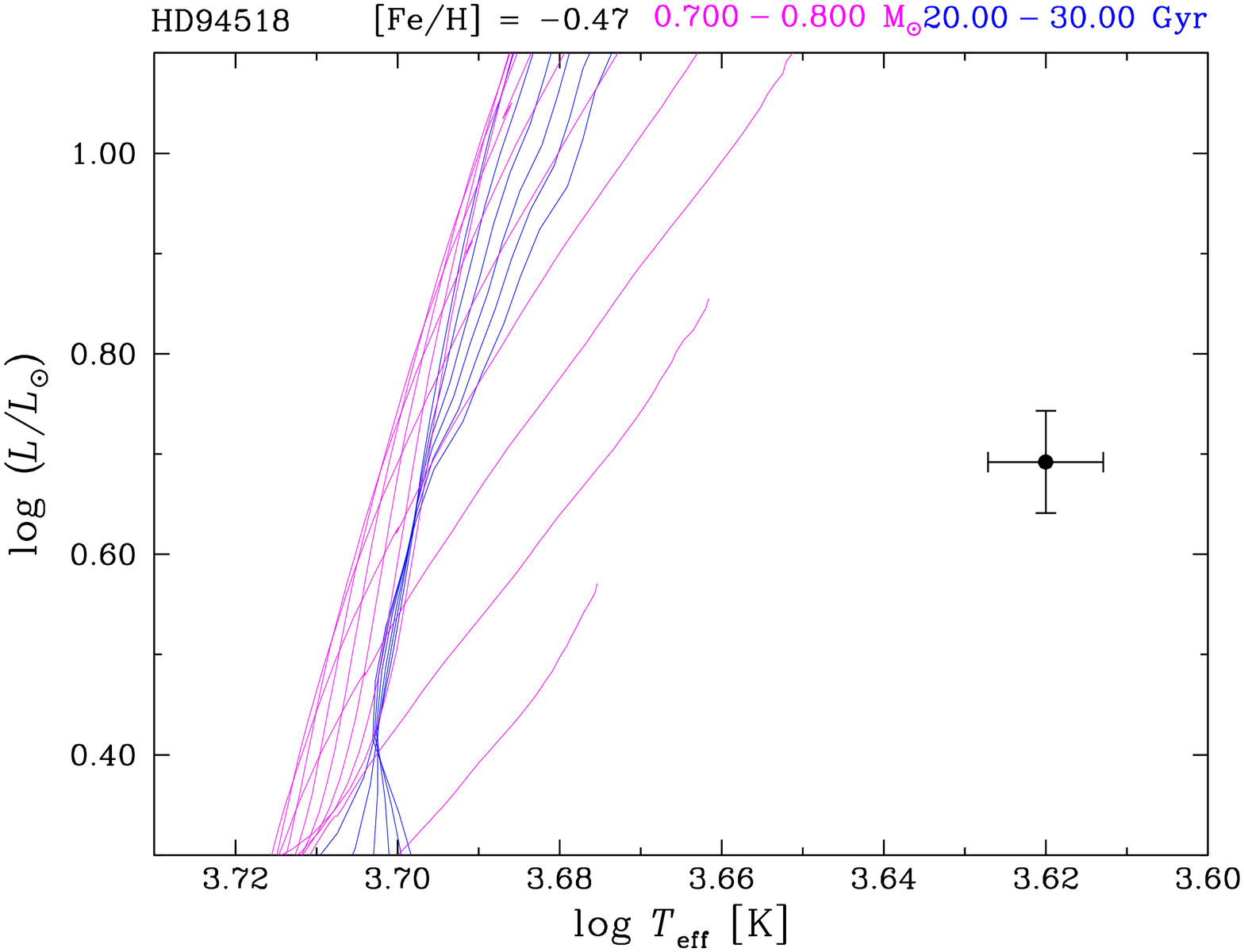}}
\end{minipage}
\caption{Evolutionary tracks (magenta lines) and isochrones (blue lines) from Yonsei-Yale, calculated for different values of metallicity, showing how we derived the stellar masses and ages. The ranges of age (increasing from left to right, \object{HD\,94518} excepted) and mass (increasing from right to left, \object{HD\,94518} excepted) of the models plotted are indicated on the top of each panel.}
\label{ev_diag}
\end{figure*}

For the volatile elements (C, N, and O), and for Na as well, we used the spectral synthesis technique, also performed with the MOOG code, to reproduce the observed spectra in regions containing molecular lines of electronic-vibrational band heads of the \C2 Swan system (centered at $\lambda$5128 and $\lambda$5165), of the CN blue system (centered at $\lambda$4215), and to atomic lines of C ($\lambda$5380.34), O ($\lambda$6300.30), and Na ($\lambda$6154.23 and $\lambda$6160.75). Again, as in Paper~1, we preferred not to use the C atomic line at $\lambda$5052.17 because it is blended with a strong Fe line, which may affect the abundance determination.

The synthetic spectra were computed in wavelength steps of 0.01~\AA, also considering the continuum opacity contribution in ranges of 0.5~\AA, line-broadening corrections (from velocity fields and instrumental broadening), and the limb darkening effect. Some parameters of atomic and molecular spectral lines were taken from the VALD database and from \citet{Kurucz1995}, respectively. Additional molecules that contribute to the spectral line formation in the studied wavelength regions are MgH (at $\lambda$5128, $\lambda$5165, and for the C atomic line), CH (at $\lambda$4215), and CN (for the Na and O atomic lines), though their contributions are relatively small. The $gf$ values of some atomic and molecular lines were revised to fit the solar spectrum, which was taken as a reference in our differential chemical analysis. The oxygen line at $\lambda$6300.30 is blended with a Ni line at $\lambda$6300.34 (\loggf\ = $-$2.3098). Therefore, we derived the O abundances by taking the Ni abundance of each star into account when fitting the observed profile. Moreover, to account for the presence of telluric lines in this region, we adopted the telluric spectrum by \citet{Wallaceetal2011}, with the resolution degraded to match our spectra.

To account for spectral line broadening from velocity fields, as in Paper~1, we adopted the same definition of \Vbroad\  (also listed in Table~\ref{spec_pars}), which is a composite of velocity fields, such as rotation and macro-turbulent velocities. The instrumental broadening was estimated from the width of thorium lines, and the linear coefficient $u$ of the stellar limb darkening was taken from \citet{DiazCordoves1995}.

Figure~\ref{stars_synth} shows the spectral synthesis applied to three of the regions used for abundance determination, the \C2\ ($\lambda$5165) and CN ($\lambda$4215) molecular band heads and the O atomic line at $\lambda$6300.30, for two dwarf stars with different spectral S/N. The synthetic spectra were  resampled in steps of 0.05~\AA\ to consistently match the wavelength scale of the observed spectra. For examples of spectral synthesis applied to other regions and to other stars, we refer the reader to Fig.~3 and 5 of Paper~1.

For a comparison between the results provided by molecular- and atomic-line C abundances, Fig.~\ref{Dcfe_teff_sn} shows the differences plotted as a function of \teff\ and of S/N for our subsamples of dwarfs, subgiants, and giants. Besides a slight underabundance (within errors, however) of molecular with respect to atomic C abundances for cool dwarfs, no systematic difference is observed in the whole range of \teff\ or S/N.

We estimated the uncertainties in the same way as in Paper~1, i.e., using our routine that includes the error propagation of the input parameters used by MOOG: the photospheric parameters \teff, \logg, [Fe/H], and $\xi$, and the broadening velocity \Vbroad. Each one in turn, the input parameters are iteratively changed by their errors, and new values of abundance are computed. The difference between new and best determination gives the error due to each parameter uncertainty, and $\sigma$([X/Fe]) is a quadratic sum of individual contributions. The error in \Vbroad\ was estimated to be of the order of 1~\kms\ or smaller.

Table~\ref{spec_pars} shows an excerpt with the derived [Ti/Fe] abundance ratios for the 140 dwarf stars. The complete tables with abundances of all the other elements, also for subdwarfs and giants, are available in electronic form at the CDS. The parameters of atomic and molecular lines are listed in Tables~5 and 6, which are only available in electronic form at the CDS as well. Table~5 contains the wavelength of the spectral feature, the atomic and molecular line identification, the lower-level excitation potential, the $gf$ values, and the dissociation energy $D_0$ (for molecular features only). Table~6 (strong atomic lines) contains the same information of Table~5, except the dissociation energy parameter.

\begin{figure*}
\centering
\resizebox{\hsize}{!}{\includegraphics{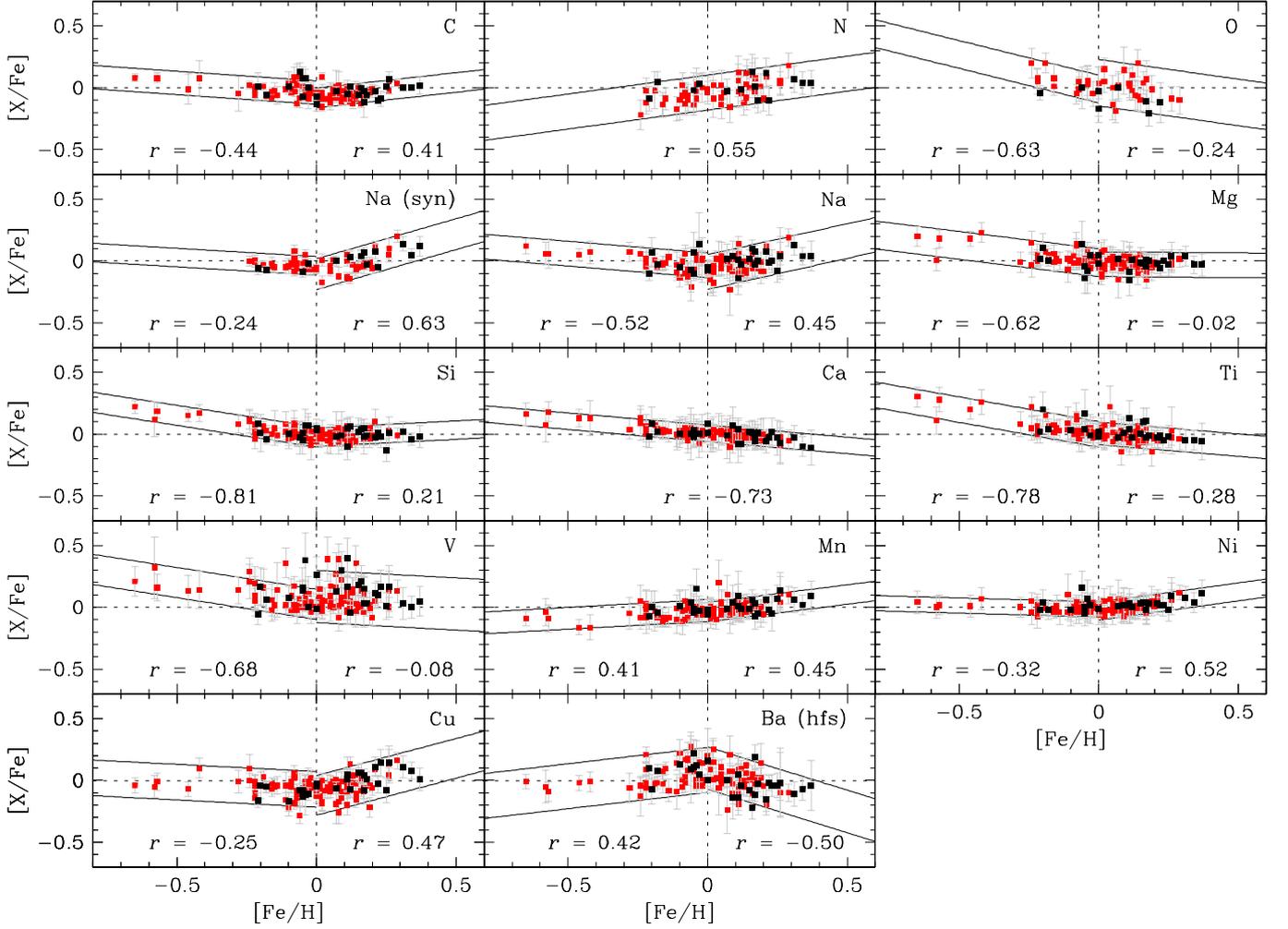}}
\caption{Abundance ratios as a function of metallicity for our sample of 120 dwarf stars of the thin disc population. Stars with planets are represented by black symbols. For sodium, the abundance ratios from both spectral synthesis and equivalent widths determination are plotted. The 95\% confidence intervals (solid lines) around linear regressions, and their respective correlation coefficients ($r$) are also shown. The linear regressions were derived either for [Fe/H] $\leq$ 0, [Fe/H] $\geq$ 0, or for the whole metallicity range, depending on which one has the more significant slope.}
\label{ab_feh_d}
\end{figure*}
\begin{figure*}
\centering
\resizebox{\hsize}{!}{\includegraphics{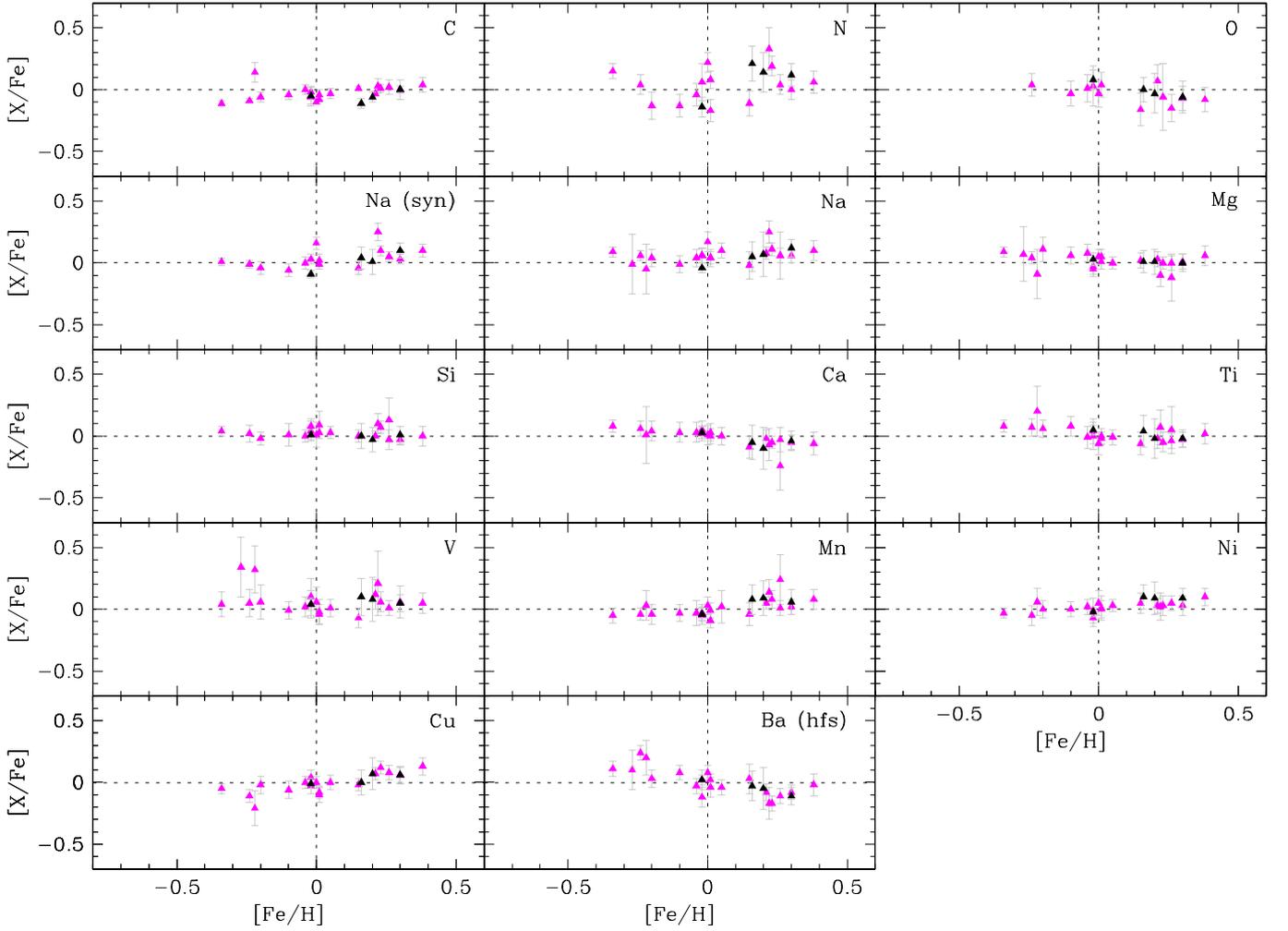}}
\caption{The same as Fig.~\ref{ab_feh_d} but showing the abundance ratios of 25 subgiant stars of the thin disc population.}
\label{ab_feh_s}
\end{figure*}
\begin{figure*}
\centering
\resizebox{\hsize}{!}{\includegraphics{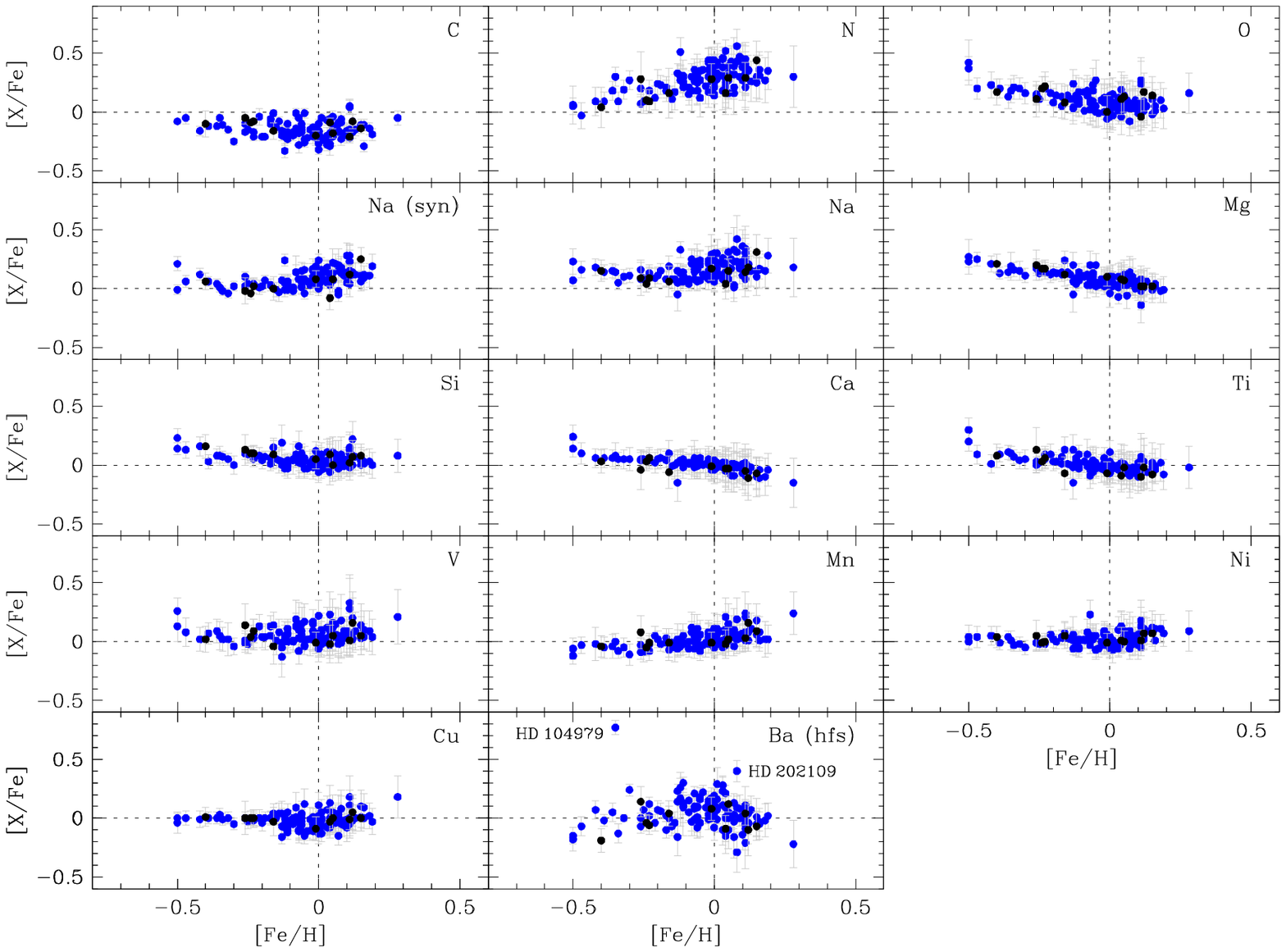}}
\caption{The same as Fig.~\ref{ab_feh_d} but showing the abundance ratios of 136 giant stars of the thin disc population. The stars indicated are discussed in Sect.\ref{outliers}.}
\label{ab_feh_g}
\end{figure*}

%
%
\section{Evolutionary and kinematic parameters}
\label{evol_kinem}

\subsection{Stellar masses and ages}
\label{mass_age}

We derived masses and ages through interpolations in the grids of Yonsei-Yale (${\rm Y}^2$) evolutionary tracks and isochrones \citep{Yietal2001,Kimetal2002} on the HR diagram. The determination of these parameters takes into account the metallicity of each star. The stars' luminosities used in the diagrams were calculated with parallaxes taken from the new reduction of the Hipparcos data \citep{vanLeeuwen2007}, and bolometric corrections ($BC$) from \citet{Flower1996}, which were revised by \citet{Torres2010}. For the Sun, we adopt $BC_{\sun} = -0.07$ and $M_{\rm bol}^{\sun} = 4.75$.

A few examples for different metallicities are illustrated in Fig.~\ref{ev_diag}. To reproduce the Sun's position in the HR diagrams, adopting \teff\ = 5777~K and age = 4.53~Gy \citep{GuentherDemarque1997}, the evolutionary tracks and isochrones were displaced in $\log(T_{\rm eff})$ by 0.001628 ($\sim$22~K in \teff) and in $\log(L/L_{\sun})$ by 0.011. By doing so, we derived for the Sun a mass $M$ = 1.003 $\pm$ 0.004~\Msun, and an age = 4.37 $\pm$ 0.17~Gyr (see Fig.~\ref{ev_diag}a). The figure also shows: a metal-poor and young star (\object{HD\,106516}, Fig.~\ref{ev_diag}b); a metal-rich one (\object{HD\,12661}, Fig.~\ref{ev_diag}c); a young star that is outside the lower border of the models (\object{HD\,26767}, Fig.~\ref{ev_diag}d); a low-mass and old star (\object{HD\,208313}, Fig.~\ref{ev_diag}e); and another star outside the limits of the models, this time in the region of low-mass and old stars (\object{HD\,94518}, Fig.~\ref{ev_diag}f).

The star \object{HD\,26767} is known to be a member of the Hyades cluster, of 625~Myr, and has been studied in several works in the past decades. Other stars either very young or outside the lower border of the models have been identified as members of the Hyades, the Ursa Major ($\sim$500~Myr), or the Pleiades (20-150~Myr) moving groups, or even classified as possible subdwarfs. In the case of old stars outside the model limits, they are likely low-mass stars with age no older than the age of the Universe \citep[13.8~Gyr,][]{Komatsuetal2011}.

Table~\ref{evol_pars} lists the derived masses and ages, together with the parameters used to compute the luminosities. The uncertainties were estimated taking into account the error propagation of the photospheric parameters and of the luminosity, using $\sigma(\pi)$ from Hipparcos, and adopting $\sigma(V)$ = 0.01 and $\sigma(BC)$ = 0.005. In view of the asymmetric distribution of the isochrones on the HR diagram, the table shows the lower and the upper uncertainties on age.

\subsection{Kinematic properties}
\label{kinem}

In the same way as in Paper~1, we grouped our stars according to their main population membership by computing the probability that a given star belongs to the Galactic thin disc, thick disc, halo, or to transition a population. First, using equations of \citet{JohnsonSoderblom1987}, we computed the space-velocity components \Ulsr, \Vlsr, and \Wlsr  with respect to the Local Standard of Rest (LSR) using parallaxes and proper motions from the new reduction of Hipparcos, and radial velocities from the ELODIE archive, from the Simbad database, and from \citet{Holmbergetal2007}. For the Sun, we adopted \Ulsr\ = 10.0, \Vlsr\ = 5.3, and \Wlsr\ = 7.2~\kms\  \citep{DehnenBinney1998}. Then, employing the equations of \citet{Reddyetal2006}, we computed the probability that a star belongs to one of the three populations. A probability $P_{\rm thin}$, $P_{\rm thick}$, or $P_{\rm halo}$ greater than or equal to 70\% classifies the star as a thin, thick, or an halo member, respectively. If the probability is in-between, the star is classified as belonging either to the thin/thick or to the thick/halo transition group.

Our sample of 309 stars contains 281 thin-disc (out of which 120 dwarfs, 25 subgiants, and 136 giants), and 16 thick-disc (10 dwarfs, 2 subgiants, and 4 giants) population members. None of them belongs to the halo, and 13 (10 dwarfs, 2 subgiants, and 1 giant) belong to a thin/thick transition population. The membership of each star is indicated in Table~\ref{evol_pars}.

%
%
\section{Results and discussion}
\label{res}

\subsection{Abundance trends with metallicity}
\label{ab_feh}

The [X/Fe] abundance ratios as a function of metallicity are plotted in Figs.~\ref{ab_feh_d}, \ref{ab_feh_s}, and \ref{ab_feh_g} for dwarfs, subgiants, and giants, respectively. Since most of our sample stars belong to the thin disc, for homogeneity, only these stars were used in the abundance analysis. They are all listed in the tables, but only thin-disc members are displayed in the figures. Besides iron, titanium is the only element in our study having spectral lines of both neutral and singly ionized atoms. We have found no significant difference between \ion{Ti}{i} and \ion{Ti}{ii}, hence we adopted an average value of all lines together.

Despite the small number of metal-poor ([Fe/H] $< -$0.3~dex) or metal-rich ([Fe/H] $>$ 0.3~dex) stars in this sample, the trends of the abundance ratios with metallicity are in good agreement with those published in the literature concerning the nucleosynthetic origin of the elements and the evolution of their abundances with time. Regarding the abundance trends for dwarfs see, e.g., \citet{Adibekyanetal2012b} and \citet{Kangetal2011}. For giant stars see, e.g., \citet{Takedaetal2008}. The abundance trends for subgiants are very similar to those for dwarfs for all the investigated elements. A few cases concerning the abundances of the three subsamples deserve particular attention, and we discuss them in the following paragraphs.

There are systematic differences in the abundance ratios of carbon, nitrogen, and sodium when the subsamples of dwarfs, subgiants, and giants are compared with each other: in the range of solar metallicity (from $-$0.1 to 0.1~dex) giant stars are underabundant in [C/Fe] by $-$0.17 $\pm$ 0.07~dex, and overabundant in [N/Fe] by 0.31 $\pm$ 0.10~dex and in [Na/Fe] by 0.16 $\pm$ 0.08~dex. The abundance trends for these elements confirm our previous result for carbon discussed in Paper~1, and agree with the results of \citet{Takedaetal2008} for sodium and carbon. Concerning oxygen, we found [O/Fe] = 0.07 $\pm$ 0.07~dex in the solar metallicity range. Therefore, we do not confirm the oxygen deficiency found by \citet{Takedaetal2008}, who used the forbidden [O\,I] line at 5577~\AA. Their oxygen abundances were in any case probably underestimated as affirmed by the authors. Our results also agree with those of \citet{Liuetal2010}, who found that [C/Fe] is depleted by 0.13~dex, [Na/Fe] is overabundant by 0.1~dex, and [O/Fe] was not altered after the first dredge-up. From a theoretical point of view, changes in the surface abundances of evolved stars is expected for C and N, and unexpected for O according to quantitative predictions made by \citet{Iben1991}.

\begin{figure*}
\centering
\resizebox{\hsize}{!}{\includegraphics{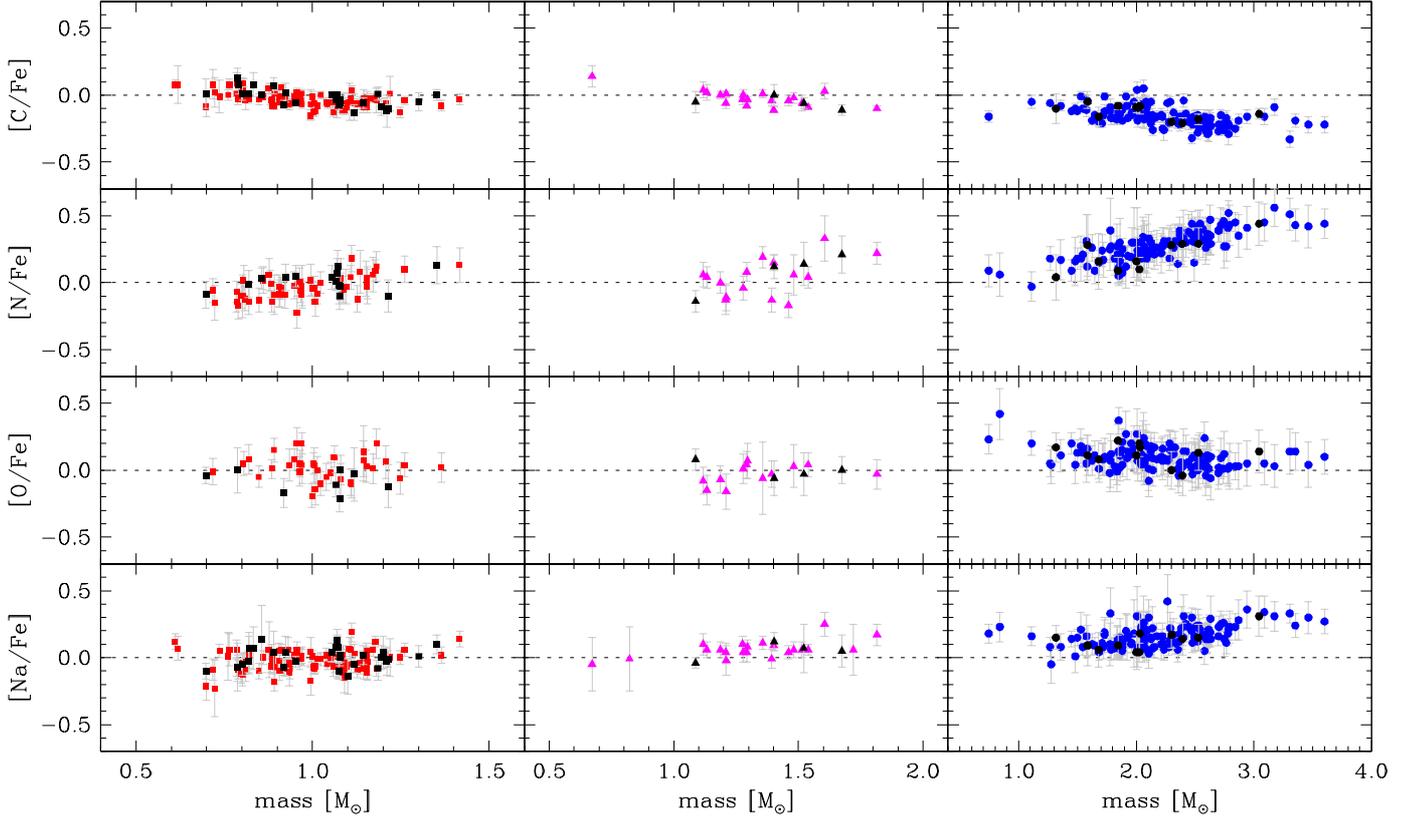}}
\caption{Abundance ratios of C, N, O, and Na as a function of mass for our sample of dwarfs (left panels), subgiants (middle panels), and giants (right panels). Stars with planets are represented by black symbols.}
\label{xfe_mass}
\end{figure*}

A commonly used explanation for such systematic differences is based on the effect of mixing processes that dredge up C-poor and N- and Na-rich material (produced by CN and NeNa cycles) to the surface of evolved stars. On the other hand, the use of spectral lines contaminated by blends is also a possibility. In order to test this, we derived the sodium abundances using the two techniques described in Sect.~\ref{abund}, from the equivalent widths and from spectral synthesis. The available Na lines in our spectral coverage are affected by ${\rm C}_2$ and CN molecular bands. Performing spectral synthesis allows the inclusion of the already derived C and N abundances and, therefore, to account for the presence of such bands. However, doing so, the [Na/Fe] overabundance in giants is only slightly reduced, not enough to cancel out the larger systematic difference with respect to dwarfs, favoring the mixing-process hypothesis. Another result that favors this hypothesis is that [C/Fe], [N/Fe], and [Na/Fe] have a trend with mass: a negative trend with increasing mass in the case of carbon, and positive ones for nitrogen and sodium (see Fig.~\ref{xfe_mass}). More massive stars have evolved faster and, therefore, have become C-poorer and N- and Na-richer than stars with masses close to the solar value. The oxygen abundances remain constant and independent of mass, supporting the above discussion.

\begin{figure*}
\centering
\resizebox{\hsize}{!}{\includegraphics{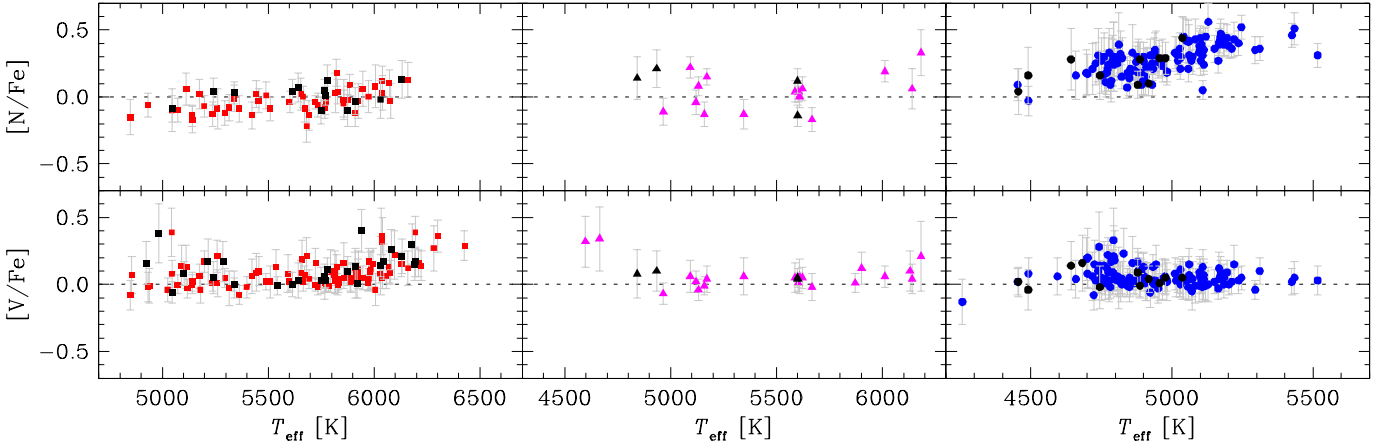}}
\caption{[N/Fe] and [V/Fe] as a function of the effective temperature for our sample of dwarfs (left panels), subgiants (middle panels), and giants (right panels). Stars with planets are represented by black symbols.}
\label{xfe_teff}
\end{figure*}

The nitrogen abundances of our giant stars show a clear dependence on the effective temperature (see Fig.~\ref{xfe_teff}, top right panel). One possibility is that this dependence is related to the continuum definition when using the CN molecular band. In this case, the reader would ask if the trend in [N/Fe] vs. [Fe/H] is also related to this dependence on \teff, due to a bad continuum definition. However, if we limit the effective temperature within the almost flat region seen in the [N/Fe] vs. \teff\ plane, from about 4600 to 5000~K, the trend of [N/Fe] with metallicity still remains. Moreover, the effective temperature of our sample of giants increases with increasing mass, i.e., it seems that more evolved stars have higher temperatures. Therefore, we believe that the [N/Fe] dependence on \teff\ is probably related to evolution effects instead of to bad continuum definition.

Concerning the carbon abundances in dwarf stars, our result is in line with those of \citet{DelgadoMenaetal2010} in the range of solar metallicities. They found, however, a steeper trend in [C/Fe] vs. [Fe/H] in the range of metal-poor stars, and practically a flat distribution of metal-rich dwarfs. On the other hand, \citet{GonzalezHernandezetal2010,GonzalezHernandezetal2013b} derived [C/Fe] ratios continuously decreasing with increasing [Fe/H] in the whole range of metallicities for their sample of dwarfs and solar analogs.

For the nitrogen abundances of dwarf stars, we derived a slope of 0.31 $\pm$ 0.06 in the [N/Fe] vs. [Fe/H] plane for dwarfs with and without planets fitted together. This is quite steeper than the one derived by \citet{Ecuvillonetal2004} for their sample of solar type stars with and without planets (0.10 $\pm$ 0.05) based on the NH band at 3360~\AA. We checked whether our [N/Fe] correlates with the effective temperature or the surface gravity, and no significant trend is observed in both cases. However, it seems that our dwarf stars with temperature higher than the solar value either have [N/Fe] close to zero or are slightly overabundant, whereas those cooler than the Sun are slightly underabundant (see Fig.~\ref{xfe_teff}, top left panel). By performing the same fit in the [N/Fe] vs. [Fe/H] plane but only to stars with \teff\ $<$ 5800~K we derive a slope of 0.28 $\pm$ 0.06, a value that is less steeper but still significantly high.

The decreasing [O/Fe] with increasing metallicity observed for our sample of dwarfs agrees with the trends derived by \citet{DelgadoMenaetal2010} and by \citet{GonzalezHernandezetal2010,GonzalezHernandezetal2013b}, and also with the recent results by \citet{BertrandeLisetal2015} for the O abundances based on the \ion{O}{i} $\lambda$6158.17 and [\ion{O}{i}] $\lambda$6300.30 atomic lines.

The distributions of [X/H] are shifted towards higher abundances (see Fig.~\ref{boxplots}) due to the metal-rich nature of stars hosting giant planets as already known and discussed in the literature \citep[see, e.g.,][]{Nevesetal2009,Adibekyanetal2012b}. The exceptions in the current study are  oxygen and barium, which [X/H] abundances show no clear enrichment in giant planet hosts (see discussion in Sect.~\ref{stat_an}). In other words, most of these elements follows the same behavior of iron and are, for some reason, also linked to the planetary formation process. It is worth noting that only stars with giant planets (mass $\gtrsim$ 50 Earth masses) were considered and compared with only stars for which no giant planet has been detected. We conducted an exhaustive search in the literature in order to classify each star and create these two subsamples. They are identified in the remarks column of Table~\ref{evol_pars}, together with the respective reference. In addition, we note that we performed this comparison only for dwarf stars of the thin disc population. Evolved stars are normally not included in the surveys of planet search, because they are well known for their intrinsic radial-velocity variation due to pulsation and/or jitter (distortion in the observed signal caused by stellar magnetic activity). Subgiant stars could be added to the subsample of dwarfs, but we preferred to keep the homogeneity.

\begin{figure*}
\centering
\resizebox{\hsize}{!}{\includegraphics{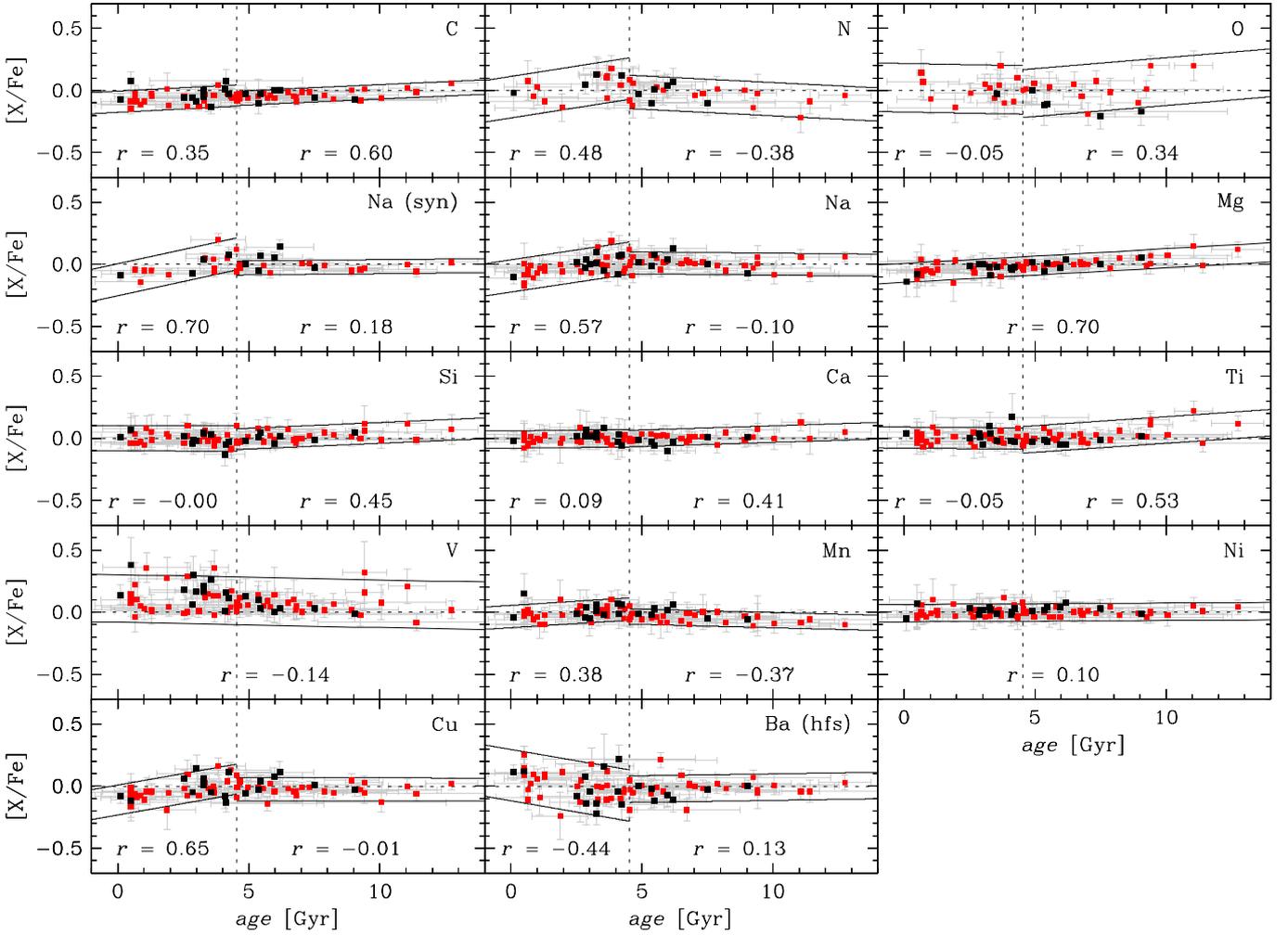}}
\caption{Abundance ratios as a function of age for our subsample of 120 dwarf stars of the thin disc population. Stars with planets are indicated by black symbols. For sodium, the abundance ratios from both equivalent widths and spectral synthesis determination are plotted. The vertical dashed line indicates the adopted solar age (4.53~Gyr). The 95\% confidence intervals (solid lines) around linear regressions, and their respective correlation coefficients ($r$) are also shown. The linear regressions were derived either for age $\leq$ 4.53~Gyr, age $\geq$ 4.53~Gyr, or for the whole range in age, depending on which one has the more significant slope.}
\label{ab_age_d}
\end{figure*}

In order to quantify the observed trends with metallicity, we fitted linear regressions in the range of stars metal-poorer or metal-richer than the Sun, or in the whole range of metallicity, depending on which one provides a higher correlation. The abundance ratios were then subtracted by the predictions of these linear fits trying to remove (or at least minimize) the effects of the chemical evolution of the Galaxy's thin disc. The abundances corrected from such effects are used and discussed in Sect.~\ref{ab_tc}, where we derive the trends with condensation temperature of the elements.

\subsubsection{Chemically peculiar stars}
\label{outliers}

Most of the dwarf and subgiant stars plotted in Figs.~\ref{ab_feh_d} and \ref{ab_feh_s} follow the abundance trends with metallicity. A particular exception seems to be the vanadium abundance ratios for some of them. Such stars have [V/Fe] higher than what is expected for their metallicity. By plotting the abundance ratios of the elements as a function of the other stellar parameters, we noticed that there is a systematic positive trend of [V/Fe] as a function of the effective temperature for \teff\ $\gtrsim$ 6000~K for our sample of dwarfs (see Fig.~\ref{xfe_teff}, bottom left panel). For this reason, in the following sections we discuss the results obtained by including or excluding the vanadium abundances of stars with temperature higher than this limit. No clear trend with temperature for the other elements is observed, and of the abundance ratios with the other stellar parameters (except age) neither. This is the case at least for dwarfs and subgiants, because the elemental abundances in giants may have been affected by stellar evolution (see discussion in Sect.~\ref{ab_feh}).

\begin{figure*}
\centering
\begin{minipage}[t]{0.365\textwidth}
\centering
\resizebox{\hsize}{!}{\includegraphics{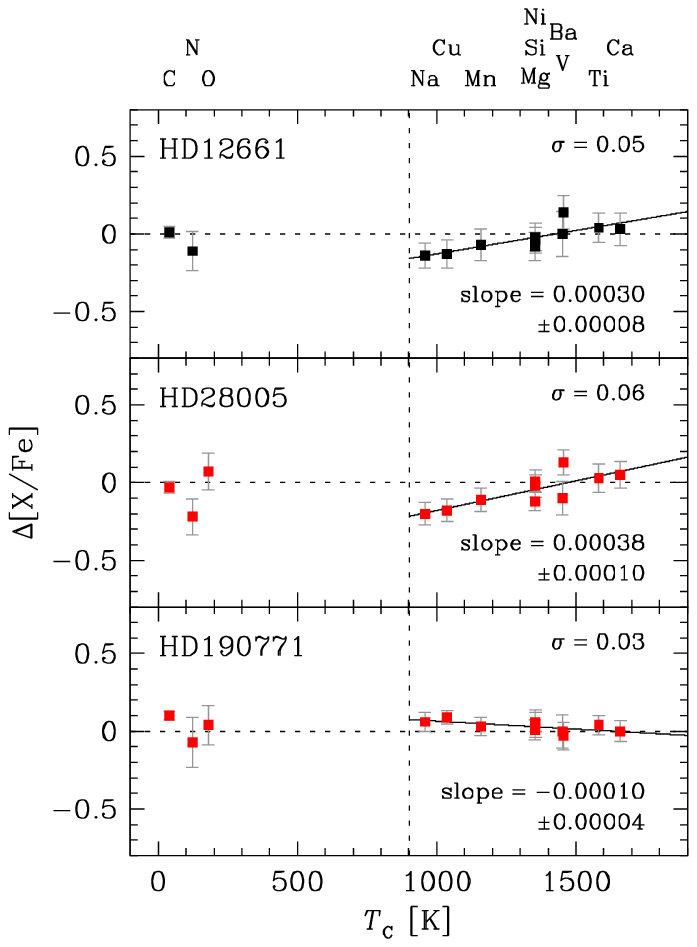}}
\end{minipage}
\begin{minipage}[t]{0.31\textwidth}
\centering
\resizebox{\hsize}{!}{\includegraphics{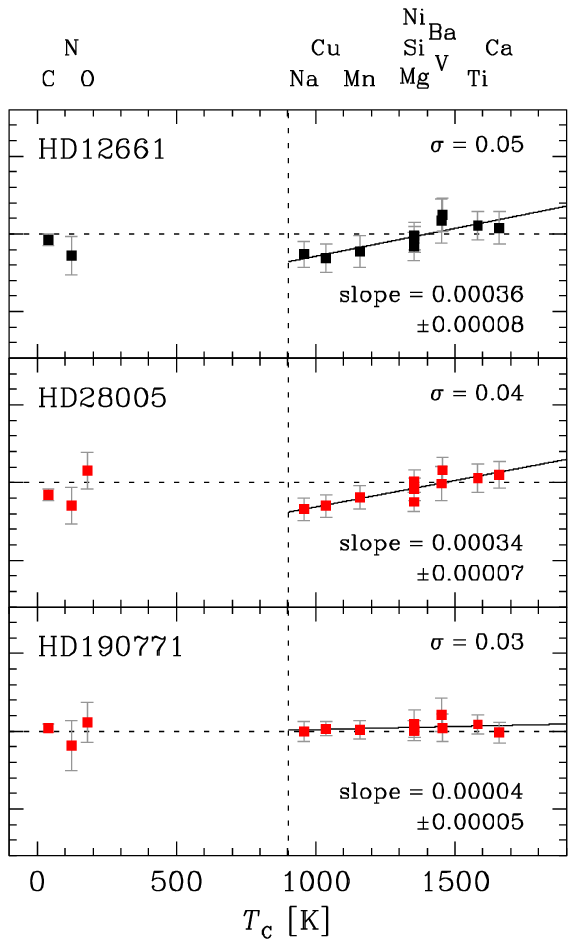}}
\end{minipage}
\begin{minipage}[t]{0.31\textwidth}
\centering
\resizebox{\hsize}{!}{\includegraphics{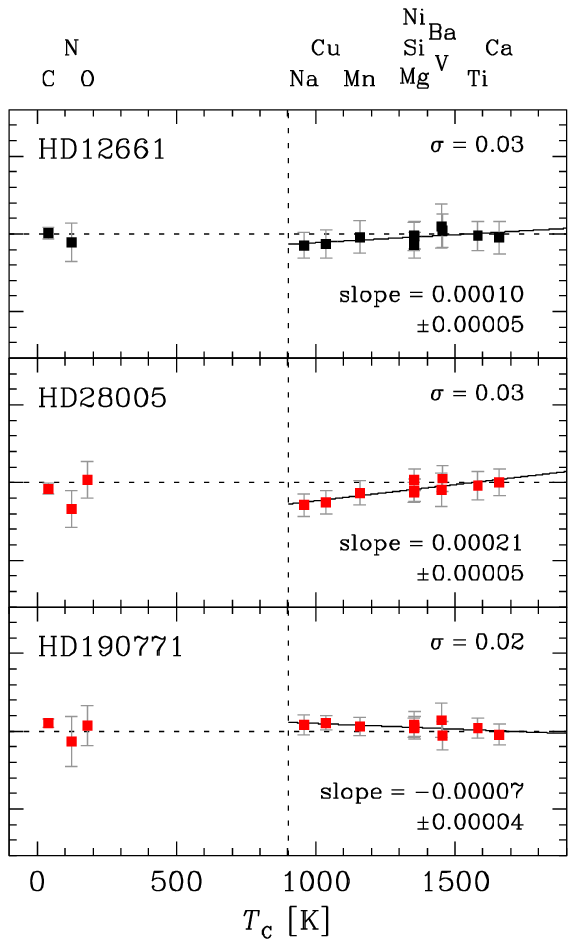}}
\end{minipage}
\caption{Abundance differences [X/Fe]$_{\rm Sun}$ $-$ [X/Fe]$_{\rm stars}$ as a function of the condensation temperature of the elements for some of our dwarf stars of the thin disc population. We show the trends before (left panels) and after applying the corrections for the Galactic chemical evolution effects based on age (middle panels) and on metallicity (right  panels). The standard deviations of the linear regressions (solid lines) for \Tc\ $>$ 900~K and their respective slopes are also shown. The planet host star is indicated by black symbols.}
\label{ab_tc_fig}
\end{figure*}

Concerning the chemical peculiarity of giant stars, two of our sample stars deserve special given their position in the [Ba/Fe] vs. [Fe/H] diagram of Fig.~\ref{ab_feh_g}. Both \object{HD\,104979} and \object{HD\,202109} appear enriched in barium compared to other giants with the same metallicity. \citet{Williams1971} derived a [Ba/Fe] = 1.1~dex for \object{HD\,104979} and first suggested it to be a \ion{Ba}{ii} star\footnote{A \ion{Ba}{ii} star originates in the following scenario: when the more massive component of a binary system evolves as a thermally pulsing asymptotic giant branch star, the material produced in the He-burning envelope, enriched in s-process elements, is dredged up to the surface and then accreted by the companion by wind mass transfer. The initially more massive star finishes as a white dwarf whereas the companion becomes a primary barium star.}. More recent works have confirmed the barium overabundance for this star, namely [Ba/Fe] = 0.93 $\pm$ 0.10~dex \citep{Zacs1994} and [Ba/Fe] = 0.54 $\pm$ 0.31~dex \citep{Smiljanicetal2007}, both in agreement with our determination of [Ba/Fe] = 0.77 $\pm$ 0.06~dex. The barium star nature of \object{HD\,202109} was first suggested by \citet{Chromeyetal1969}, who measured [Ba/Fe] = 0.8 $\pm$ 0.3~dex, and afterward classified as mild (marginal enriched) barium stars: [Ba/Fe] = 0.41 $\pm$ 0.13~dex by \citet{Zacs1994}, and [Ba/Fe] = 0.31 $\pm$ 0.26~dex by \citet{Smiljanicetal2007}. We derived [Ba/Fe] = 0.40 $\pm$ 0.09~dex, a result that agree with the mild barium star classification.

\subsection{Abundance trends with age}
\label{ab_age}

Figure~\ref{ab_age_d} shows the [X/Fe] abundance ratios as a function of age for the subsample of dwarf stars of the thin disc population. Similarly to what is described in Sect.~\ref{ab_feh} in the case of the metallicity, we also quantified the observed trends with age by fitting linear regressions in the range of stars younger or older than the Sun, or in the whole range of age, depending on which one provides a higher correlation. Again, the abundance ratios were subtracted by the predictions of these linear fits in order to remove (or at least minimize), using another approach, the effects of the chemical evolution of the Galaxy. The abundances corrected from such effects are used and discussed in Sect.~\ref{ab_tc}, where we derive the trends with condensation temperature of the elements.

Depending on the position of the star in the HR diagram, the errors in age may be quite high, and the age determination is thus less reliable. By means of the distribution of the uncertainties on this parameter, we set a 2$\sigma$ confidence limit up to which about 95\% of the stars are found to have errors smaller than that limit, namely $\sim$4~Gyr. Hence, only stars with $\sigma$(age) $<$ 4~Gyr was considered in the analysis.

\subsection{Abundance trends with condensation temperature}
\label{ab_tc}

During the last few years, several works have investigated the relation between the stellar chemical abundances and the condensation temperature (\Tc) of the elements in stars with and without planets. Aiming to minimize systematic errors, the abundance difference between the Sun and other stars, $\Delta$[X/Fe], has been used, where the solar abundances are derived from, e.g., a Moon's spectrum.

\begin{figure*}
\centering
\resizebox{\hsize}{!}{\includegraphics[angle=-90]{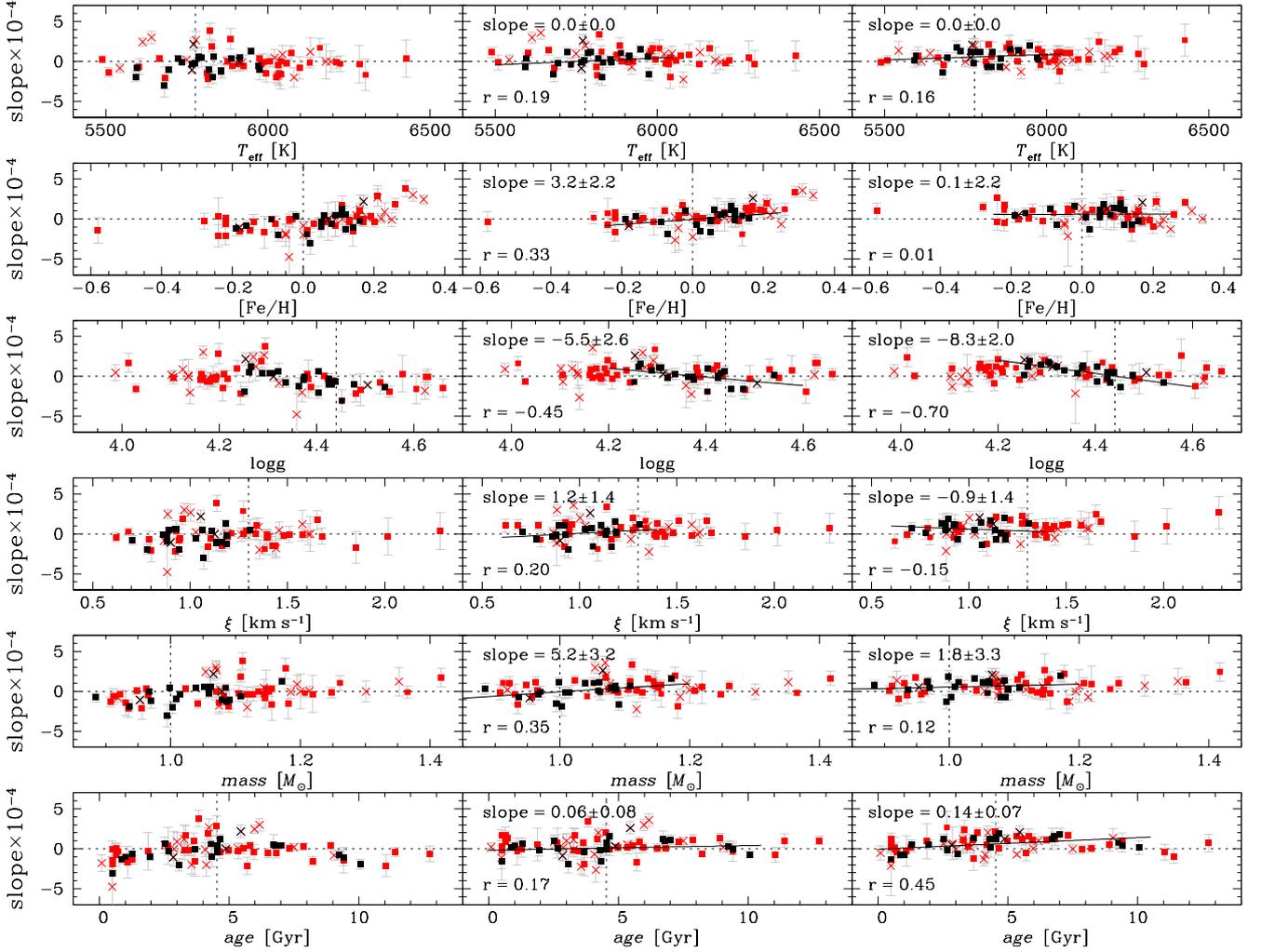}}
\caption{Slope of the regressions computed in the diagrams $\Delta$[X/Fe] vs. \Tc\ (Fig.~\ref{ab_tc_fig}) plotted as a function of the derived stellar parameters. We show possible trends before (left panels) and after applying the corrections for the Galactic chemical evolution effects based on age (middle panels) and on metallicity (right panels). Dwarf stars of the thin disc population are separated in solar analogs (red symbols) and non-solar analogs (black symbols). Stars with planets are indicated by crosses. The slopes of linear regressions (solid lines) applied to the solar analogs and their respective correlation coefficients ($r$) are also shown.}
\label{ab_slope}
\end{figure*}

On the one hand, it has been suggested that the observed trends of $\Delta$[X/Fe] as a function of \Tc\ may have some relation with the planetary formation, in particular, with the formation of terrestrial planets \citep[see the review performed by][]{Ramirezetal2010}. On the other hand, other authors argued that this kind of relation practically vanishes after removing the Galactic chemical evolution effects \citep[see][and references therein]{GonzalezHernandezetal2013b}. Indeed, positive slopes in these diagrams may be the result of having higher abundances of less refractory elements, such as Na and Cu, with respect to more refractory ones, such as Ti and Ca. In Fig.~\ref{ab_feh_d}, both [Ti/Fe] and [Ca/Fe] have a slight decrease with increasing metallicity whereas [Na/Fe] and [Cu/Fe] clearly increase for higher metallicities.

In particular, some of the studies discussed in the papers above consider only solar twins (\teff\ = 5777 $\pm$ 100~K, [Fe/H] = 0.0 $\pm$ 0.1~dex, \logg\ = 4.44 $\pm$ 0.10~dex) or solar analogs (\teff\ = 5777 $\pm$ 200~K, [Fe/H] = 0.0 $\pm$ 0.2~dex, \logg\ = 4.44 $\pm$ 0.20~dex) in their analysis. Recently, \citet{Adibekyanetal2014} have published the results of their study in which they searched for correlations of the $\Delta$[X/Fe] vs. \Tc\  trends with some stellar and orbital parameters. After correcting the abundances from the Galactic chemical evolution effects, they found that the observed trends correlate with age and anticorrelate with the surface gravity, and possibly with the mean Galactocentric distance, and that the correlations (or anticorrelations) are even steeper and more evident if only solar analogs are plotted.

Here we also investigate the relations involving $\Delta$[X/Fe] as a function of the condensation temperature of the elements (to compute [X/Fe]$_{\rm Sun}$, we used the solar abundances derived from the spectrum of the sunlight reflected by the Moon). First, we fitted linear regressions for elements with \Tc\ $>$ 900~K, a limit normally used to distinguish refractory from volatiles. Values of \Tc\ were taken from \citet{Loddersetal2009}. In order to account for any effects that the Galactic chemical evolution may have on the observed trends, we applied the corrections based on both metallicity and age, as described previously. Figure~\ref{ab_tc_fig} shows a few examples. Table~8, available in electronic form at the CDS, lists the slopes of the linear regressions and the respective standard deviations for all dwarfs. Only dwarf stars were used in this analysis because their abundances are normally not affected by
stellar evolution. The table contains the slopes and their uncertainties, the correlation coefficients, and the standard deviations of the regressions before and after applying the corrections for the Galactic chemical evolution effects based on age and on metallicity.

We then plot (Fig.~\ref{ab_slope}) the respective slopes as a function of the stellar parameters \teff, [Fe/H], \logg, $\xi$, mass, and age. In this figure, solar analogs are distinguished from non-solar analogs, and stars with planets from those without planets. Regarding the sample of solar analogs, the figure shows:
$i)$ a marginal correlation with metallicity (in the sense that metal-richer stars have more positive slopes) and with mass after removing the effects of chemical evolution based on age; however, these correlations completely disappear if the correction is based on metallicity;
$ii)$ a clear anticorrelation with the surface gravity in both scenarios of Galactic chemical evolution correction;
$iii)$ a marginal correlation with age, slightly steeper after applying the correction based on metallicity.
The directions of these correlations agree with those reported by \citet{Adibekyanetal2014}. We cannot check the \Tc\ trends with Galactocentric distance because we only have solar neighborhood stars (distance to the Sun $<$ 100~pc). Concerning the relation with the presence of planets, the small number of planet hosts in our sample of solar analogs prevents us from any conclusions.

By excluding from the analysis the vanadium abundances of stars with \teff\ $\gtrsim$ 6000~K, the correlations and anticorrelations obtained above are still the same, specially because this limit is already imposed when only solar analogs are used.

%
\section{Statistical analysis}
\label{stat_an}

\begin{figure*}
\centering
\includegraphics[width=0.9\textwidth]{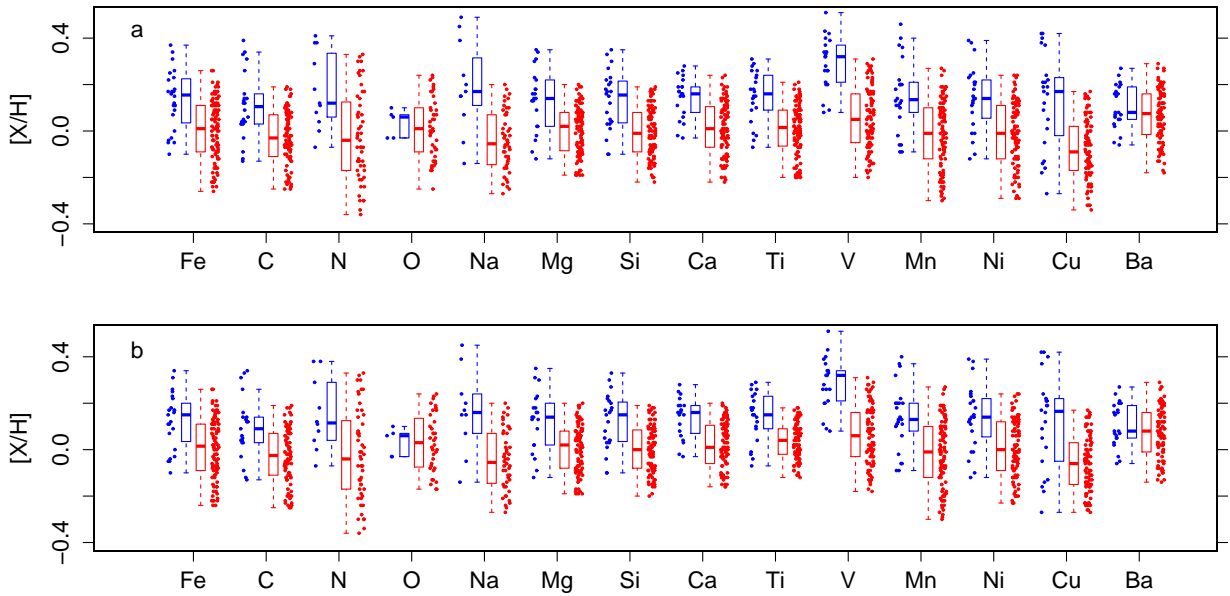}
\caption{Boxplot for the comparison between the elemental abundances of dwarf stars with and without planets, show in blue and red, 	 respectively. Two sigma clipping was applied to all abundance 	 distributions before the comparison, to avoid abundance outliers. The top panel shows the distributions for dwarfs in the [Fe/H] abundance space, whereas the bottom panel restrict the comparison to those stars having $-$0.25 $\le$ [Fe/H] $<$ +0.35. The original data points are shown beside each corresponding box to allow the visualization of the intrinsic scatter.}
\label{boxplots}
\end{figure*}

We have used a number of statistical methods \citep[see][]{Breimanetal1984} to explore the multivariate properties of our sample. We are interested in uncovering significant differences in the stellar properties for stars with and without a known giant planet. Hereafter we will use the indices 1 and 0 to distinguish these two samples, respectively. Moreover, the sample of dwarf stars will be prefixed by the letter ``d''. Thus, samples d1 and d0 stand for dwarfs with and without a known giant planet, respectively. The analysis and the following discussion is restricted to the dwarfs because their sample is thought to be less affected by an observation bias.

\subsection{Abundance distributions}
\label{ab_dist}

We compare the abundance distributions between the samples d0 (red) and d1 (blue) in Fig.~\ref{boxplots}. The boxes and whiskers summarize graphically percentiles of these abundance distributions. Particularly, the middle line inside each box marks the median for that distribution, whereas the upper and lower edge of the box indicate the interquartile range. We applied a 2$\sigma$ clipping in the abundance data of each single element in order to get rid of outliers in the abundance space. Panel (a) shows the comparison for all dwarf stars after applying the sigma clipping, while panel (b) restricts the comparison to those stars having $-$0.25 $\le$ [Fe/H] $<$ +0.35 (a range in common between the subsamples of stars with and without planets). Both comparisons depict that dwarf stars with giants planets (d1) have generally larger abundances in nearly all elements compared to dwarfs without giant planets (d0). We note that vanadium seems to be exceptionally abundant in d1 stars, whereas no significant difference is seen between d1 and d0 for the oxygen and barium abundances. Concerning vanadium, the overabundance of d1 with respect to d0 does not change if only stars with \teff\ $\lesssim$ 6000~K are used.

We used the Welsch's t-test to quantitatively assess whether the means of these distributions are significantly different from one another. Table~\ref{ttest_lda} gives the test statistics for stars having $-$0.25 $\le$ [Fe/H] $<$ +0.35, after applying 2$\sigma$ clipping. Some $p$-values are very low ($< 0.001$), reinforcing the visual comparison provided by Fig.~\ref{boxplots}. The mean of the abundance distribution of all elements, but O and Ba, are higher in d1 stars, compared to d0 stars. Oxygen abundances are not very well represented in our sample, and only 32 d0 and 4 d1 stars have measured O abundance; this could be one of the reasons for why the t-test gave a large $p$-value for the O abundance distribution comparison. Notwithstanding, it is highly interesting, although not clear, why Ba is not particularly more abundant in d1 stars.

\begin{figure*}
\centering
\includegraphics[width=0.8\textwidth]{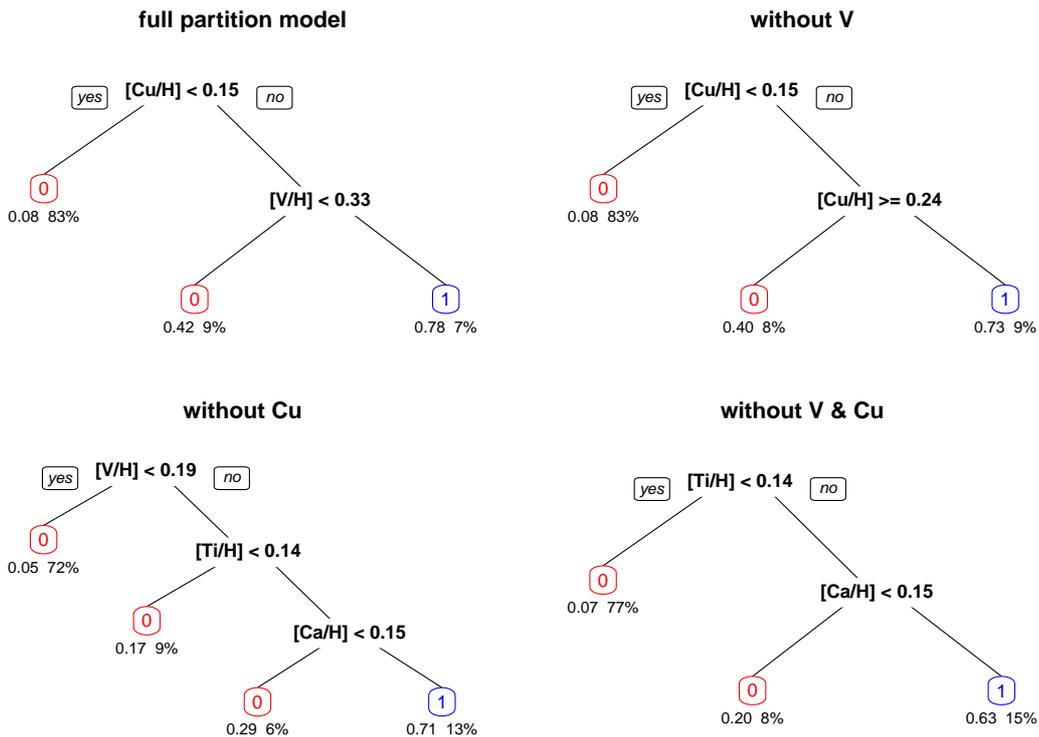}
\caption{Classification trees for the planet-harboring status amongst dwarf          stars in our sample. The nodes are annotated with a categorical binary variable that assumes values 1 or 0 for stars with and without a giant planet, respectively. The leaves are marked with statistics of the partitioned group: fraction of d1 stars in the group (stars harboring a giant planets) and percentage of all dwarfs stars that were classified in that leaf. The top left tree is the full partition model for our sample. The other trees consider the case when two important elements are not known: V and Cu, one at a time or both.}
\label{rpartmod}
\end{figure*}

The t-tests described above can only be applied to the abundance distribution of a single element at a time. They can pinpoint elements that are particularly more or less abundant in one of the tested distributions. To use the information encoded in the whole abundance space, we also applied a multivariate analysis of variance (MANOVA) to assess for the significance of differences in the multivariate means (that is, the location of samples 0 and 1 in the abundance space). According to this MANOVA, we can reject the null hypothesis that the d0 and d1 samples have similar locations in the abundance space ($p = 0.006$).

\subsection{Abundance difference per metallicity bin}
\label{z_value}

We have also considered a different approach to disentangle the effects of the chemical evolution of the Galaxy from the metallicity--giant planet connection. We have defined a crude abundance quantity $\cal Z$ = [Mg/H] + [Si/H] + [Ca/H] + [Ti/H] + [V/H] + [Mn/H] + [Ni/H] + [Cu/H] + [Ba/H]. $\cal Z$ does not have an exact physical meaning, on account of the sum of unscaled log abundances, but it can serve as enrichment index for the purpose of this analysis. For instance, it is clear that $\cal Z$ and [Fe/H] should be correlated as a result of the Galactic chemical evolution. This index could only be defined for those stars having a complete set of bundances of the elements Mg to Ba, including Fe. CNO and Na abundances were not included in the $\cal Z$ definition because this would lower substantially the final number of stars having a complete set of abundance measurements.

The index was used in a linear model of the type
\begin{equation}
{\cal Z} = \hat\beta_0 + \hat\beta_1{\rm [Fe/H]} + \epsilon,
\label{calZ}
\end{equation}
fitted by least squares to the d0 stars. The residuals of the fit are given by $\epsilon$ for each star. Following this, we calculated the residuals of the d1 stars with respect to the same linear model fitted to the d0 stars. A Welch's t-test applied to these residual distributions indicate that the difference in their means are significant (a $p$-value $p$ = 0.0023), with $\bar\epsilon_{\rm d1} - \bar\epsilon_{\rm d0} = 0.16 \pm 0.10$. That is, for a given [Fe/H], d1 stars have generally larger $\cal Z$ indices than d0 stars. If we can consider the d0 stars as a ``null hypothesis'' population, for which the elemental abundances only reflect the Galactic chemical evolution, the residuals of $\cal Z$ for the d1 stars using the linear model in Eq.~\ref{calZ} should have no chemical evolution bias. Again, if we calculate $\cal Z$ excluding [V/H], this result practically does not change (the new $p$-value would be $p$ = 0.0026). On the other hand, if $\cal Z$ is calculated excluding the [Ba/H] abundance ratios, we obtain $p$ = 0.0005 and $\bar\epsilon_{\rm d1} - \bar\epsilon_{\rm d0} = 0.22 \pm 0.12$, i.e., the difference between the d0 and d1 populations is even larger.

We defined $\cal Z$ by mixing elements with different nucleosynthetic origins, which might veil the real difference in abundance between stars with and without planets. However, the mean [X/H] distributions for $\alpha$ (Mg, Si, Ca, Ti) and for iron-peak (V, Cr, Mn, Co, Ni) elements in stars hosting giant planets are very similar, as reported by \citet{MataSanchezetal2014}. Indeed, if we calculate $\cal Z$ for $\alpha$ elements only we obtain $\bar\epsilon_{\rm d1} - \bar\epsilon_{\rm d0} = 0.08 \pm 0.06$, whereas for our iron-peak elements (V, Mn, Ni) only we have $\bar\epsilon_{\rm d1} - \bar\epsilon_{\rm d0} = 0.10 \pm 0.06$. For stars without planets, the [X/H] distributions are also very similar, having a maximum abundance at $\sim$$-$0.1 for most species \citep{MataSanchezetal2014}. Therefore, we conclude that the mix of elements when calculating $\cal Z$ is not a bad approach, specially because the differences that we are comparing become more significant when the abundances of all these elements are added together.

\subsection{Classification trees}
\label{class_trees}

Assuming the general connection between chemical abundance and the presence of a giant planet, we can extend the reasoning in using the iron abundance as an indicator of the plausibility that the star harbors a planet. To do this, we applied a classification tree to a complete set of [X/H] abundances for the dwarfs in our sample. Classification trees give the optimal splitting of observations coming from different classes (or populations). In our problem, we consider two classes: stars with and without giant planets. The tree is built through recursive partitioning of the abundance space into nested groups, aiming at terminal groups having highest purity (that is, having most, if not all, of its members coming from a single class). Each classification tree has nodes, branches and terminal leaves (or group). The nodes are annotated with some property (or variable) and a splitting rule which points to a branch or the other. The leaves are usually marked with statistics of the partitioned group.

As previously explained, on account of the requirement to have a complete set of abundances, CNO and Na measurements were not used in this analysis. No sigma clipping to the abundances was used before applying the classification trees. We show in Fig.~\ref{rpartmod} four classification trees. The leaves are marked (and color-coded, for easy identification) with the most likely class of its members: 0 (red) if the star is not likely to harbor a giant planet, and 1 (blue) if the opposite holds. Two numbers are annotated bellow each terminal leaves, showing the fraction of d1 stars in the group (stars harboring a giant planet) and the percentage of all dwarfs stars that were classified in that leaf. Thus, the numbers 0.42~9\% can be read as 9\% of all dwarfs were classified in that group and 42\% of them harbors a giant planet. The upper left panel gives the classification tree when we consider all elements from Mg to Ba. Although this abundance space is formed by the measurements of 10 chemical species, only two of them are needed for an optimal partition: Cu and V. Copper abundances seems to be the best single indicator of the presence of a giant planet: only 8\% of the stars having [Cu/H] $<$ 0.15 dex harbors a giant planet. Vanadium comes second in refining this: 78\% of stars having [Cu/H] $\ge$ 0.15 dex and [V/H] $\ge$ 0.33 dex harbors a giant planet. The three other classification trees in Fig.~\ref{rpartmod} shows the partition model when Cu, V or both variables are not used. If vanadium is taken out of the independent variable set, the whole 10-dimensional abundance space could still be partitioned using the Cu abundances alone. The purity is somewhat lower: 73\% of the stars having $0.15 \le {\rm [Cu/H]} \le 0.24$ harbors a giant planet. If, on the other hand, we take Cu from the set of independent variables, V takes the lead as the most relevant element, followed by two more splittings using Ti and Ca. The two more meaningful (extreme) leaves give: $i)$ 5\% of stars having [V/H] $< 0.19$ dex harbors a giant planet; and $ii)$ 71\% of stars having simultaneously [V/H] $\ge 0.19$, [Ti/H] $\ge 0.14$ and [Ca/H] $\ge 0.15$ dex harbors a giant planet. Taking both Cu and V out of the set of independent variables yields a classification tree based on Ti and Ca abundances that is an exact subtree from the third case (with no Cu abundances). 

\subsection{Linear discriminant analysis}
\label{lda}

\begin{table}
\centering
\caption{Probabilities of the t-test statistics and coefficients of the linear discriminant for the d0 and d1 classes.} 
\label{ttest_lda}     
\centering                                      
\begin{tabular}{cccc r@{}l}          
\noalign{\smallskip}\hline\hline\noalign{\smallskip}
\parbox[c]{1.5cm}{\centering Abundance ratio} &
$N({\rm d0})$ & $N({\rm d1})$ & $p$-value &
\multicolumn{2}{c}{\parbox[c]{1.6cm}{\centering Linear discriminant}} \\
\noalign{\smallskip}\hline\noalign{\smallskip}
 {[Fe/H]} & 94 & 23 & $2.39\times 10^{-4}$ & \,\,\,\, $-$0&.87    \\
 {[C/H]}  & 86 & 21 & $5.12\times 10^{-4}$ & \,\,\,\,	  &\ldots \\
 {[N/H]}  & 51 & 10 & $6.82\times 10^{-3}$ & \,\,\,\,	  &\ldots \\
 {[O/H]}  & 36 &  5 & $9.21\times 10^{-1}$ & \,\,\,\,	  &\ldots \\	      
 {[Na/H]} & 48 & 10 & $4.92\times 10^{-3}$ & \,\,\,\,	  &\ldots \\
 {[Mg/H]} & 89 & 20 & $8.38\times 10^{-4}$ & \,\,\,\, $-$5&.43    \\
 {[Si/H]} & 87 & 23 & $1.43\times 10^{-4}$ & \,\,\,\,	 0&.70    \\
 {[Ca/H]} & 88 & 20 & $7.62\times 10^{-6}$ & \,\,\,\, $-$0&.14    \\
 {[Ti/V]} & 79 & 21 & $2.76\times 10^{-4}$ & \,\,\,\,	 8&.23    \\
 {[V/H]}  & 85 & 21 & $2.03\times 10^{-9}$ & \,\,\,\,	 2&.27    \\
 {[Mn/H]} & 91 & 23 & $3.56\times 10^{-5}$ & \,\,\,\, $-$3&.50    \\
 {[Ni/H]} & 88 & 23 & $1.64\times 10^{-4}$ & \,\,\,\,	 5&.62    \\
 {[Cu/H]} & 81 & 22 & $6.76\times 10^{-4}$ & \,\,\,\,	 0&.99    \\
 {[Ba/H]} & 82 & 22 & $1.91\times 10^{-1}$ & \,\,\,\, $-$2&.01    \\
\hline                                             
\end{tabular}
\tablefoot{The second and third columns gives the number of stars in each group. The $p$-value gives the probability that the mean of both d0 and d1 distributions are statistically similar.}
\end{table}

A similar, but more sophisticated, approach can be done using a linear discriminant analysis (LDA), which finds the linear combination of observed properties (in our case, abundances) which best separates two or more classes of objects. The result is an index, or system of indices, that may act as a linear classifier. The difference from the classification tree is that this one uses a hierarchical recursive splitting of single properties, aiming at classes with highest purity possible, while the LDA searches for rotations of the original multi-parametric space that leads to the best discrimination of the classes in the entire space.

\begin{figure}
\centering
\includegraphics[width=\columnwidth]{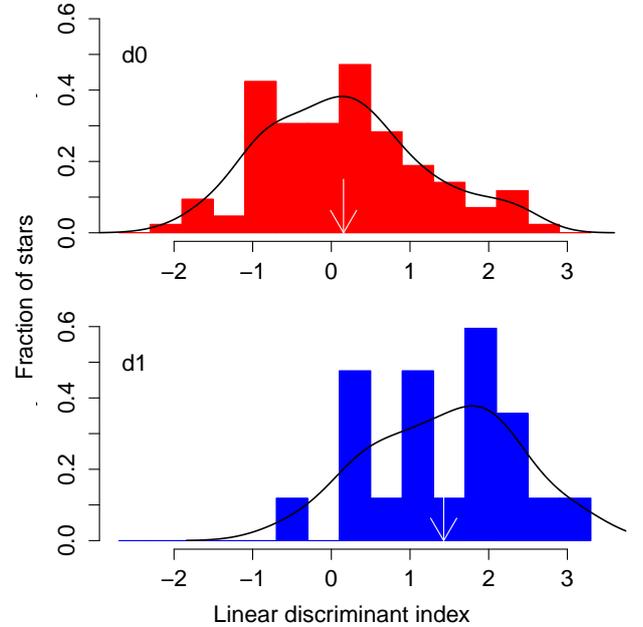}
\caption{Histograms of the linear discriminant indices defined in          Table~\ref{ttest_lda}. The color code blue and red is used to indicate stars with (top panel) and without planets (bottom panel), respectively. A black line shows the kernel density function for these indices, and a white arrow marks the position of the mean in each distribution.}
\label{LDAplot}
\end{figure}

We applied the LDA to the complete abundance set of Mg to Ba for dwarfs, looking for the linear combination of abundances that could better discriminate the d0 from d1 samples. The resulting linear discriminant is given in the last column of Table~\ref{ttest_lda} and in Fig.~\ref{LDAplot}. As explained, the LDA finds the best linear combination of the original abundances that discriminate amongst the tested classes. The numbers in the table are the coefficients of this linear discriminant. An index formed by $-0.87{\rm [Fe/H]}-5.43{\rm [Mg/H]}+\ldots+0.99{\rm [Cu/H]} -2.01{\rm [Ba/H]}$ can thus be used to discriminate d0 from d1 stars. If our sample can be taken as unbiased and representative of the stellar abundances of stars with and without giant planets, we can use this discriminant index in a predictive way, when the giant planet-harboring status of the star is unknown a priori. As pointed before, the concept is similar to using a classification tree, although now we use all abundances to build a single optimizer classifier. The power of this classifier can be seen in Fig.~\ref{LDAplot}, where we show the histogram of the discriminant index. 

We explored all possible combinations of abundance ratios involving the elements from Mg to Ba, using the Welch's test to assess interesting pairs. After considering all possibilities, we have found that [V/Ca] seems to be a good abundance ratio for discriminating dwarfs with planets: the average difference in [V/Ca] between the d0 and d1 groups amounts to 0.07 dex, leading to the rejection of the null hypothesis with $p=0.005$.

\begin{figure}
\centering
\includegraphics[width=\columnwidth]{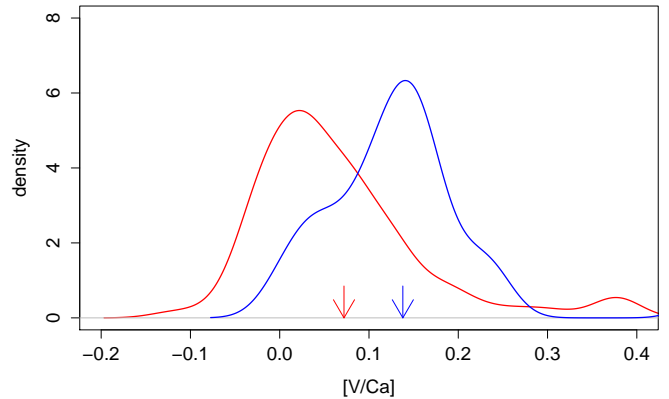}
\caption{Kernel probability density functions for the relevant abundance ratio, [V/Ca], that allows the discrimination of dwarfs with and without giant planets. The color code blue and red is used to indicate stars with and without planets, respectively. These abundance ratio was found to discriminate dwarfs with respect to their planet-harboring status, in a similar role to that played by the linear discriminant indices.}
\label{AbundRatioLDA}
\end{figure}

Kernel probability density functions for the relevant abundance ratio that allows the discrimination of dwarfs with and without giant planets are shown in Fig.~\ref{AbundRatioLDA}. The color code is similar  to that used before: blue and red for stars with and without planets, respectively. The figure shows the [V/Ca] amongst dwarf stars. Although there is  considerable intersection between the red and blue curves, it is both remarkable and puzzling that this abundance ratio seems to discriminate stars according to their planet-harboring status.

%
%
\section{Conclusions}
\label{concl}

The inclusion in the current study of more stars, with and without detected planets, and more elements aimed to improve the work presented in Paper~1. A summary of our new results and conclusions is as follows:

\begin{itemize}

\item[\it i)] the systematic differences in the abundance ratios of C, N, and Na among dwarfs, subgiants, and giants seem to confirm that mixing processes, together with cycles of nucleosynthetic reactions, do modify the abundances in the surface of evolved stars;

\item[\it ii)] such mixing processes seem, indeed, to alter the Na photospheric abundances in giant stars considering that, even tanking into account the influence of \C2 and CN molecular features, a Na overabundance is still observed;

\item[\it iii)] the slopes in the $\Delta$[X/Fe] vs. \Tc\ diagrams show a correlation with age and an anticorrelation with the surface gravity, even after accounting for the effects of the Galactic chemical evolution; these results appear only when the sample is restricted to solar-analog stars; in other words, as also pointed out by \citet{Adibekyanetal2014}, older and more evolved stars are, in same way, connected to a lower amount of more refractory elements available for the star formation; no significant correlation with other stellar parameters, and no relation with the presence of planets are observed;

\item[\it iv)] our statistical analysis has provided the following results:

$a)$ the overabundance of other elements in dwarf stars hosting giant planets confirms the previous statement that not only iron is linked to the planetary formation process, but also C, N, Na, Mg, Si, Ca, Ti, V, Mn, Ni, and Cu;

$b)$ for O and Ba there are no significant differences between the samples of dwarfs with and without giant planets; oxygen is not very well represented in our analysis, but the result for barium is intriguing and deserves further investigation;

$c)$ vanadium is the element having the most significant abundance difference between giant planet hosts and single dwarfs, and this difference is still significant if only stars with \teff\ $\lesssim$ 6000~K are used;

$d)$ by combining together the elements from Mg to Cu according to Eq.~\ref{calZ}, we found a significant overabundance, for the same metallicity, in giant planet hosts in comparison with our sample of single dwarfs;

$e)$ based on the classification tree analysis we have found that abundances of copper and calcium are special discriminators of dwarfs with and without giant planets (over the [X/H] scale);

$f)$ by means of a linear discriminant analysis, we have found that [V/Ca] seems to be a good abundance ratio for discriminating dwarfs with giant planets.

\end{itemize}

\begin{acknowledgements}
R. da Silva thanks the Instituto Nacional de Pesquisas Espaciais (INPE) for its financial support in the form of a grant (PCI-DA 300.422/2011-3, MCTI/INPE/CNPq). We thanks Jorge Mel\'endez for critically reading the manuscript. This research has made use of the SIMBAD database, operated at CDS, Strasbourg, France.
\end{acknowledgements}

\bibliographystyle{aa}
\bibliography{daSilvaetal2015}

\clearpage

\begin{table*}[p]
\centering
\caption[]{Atomic line parameters of the elements used in the analysis and            $EW$s measured in the degraded spectrum of the Solar Flux Atlas.}
\label{linelist}
\begin{tabular}{c c c r@{}l r@{}l | c c c r@{}l r@{}l | c c c r@{}l r@{}l}
\hline\hline
 & & & & & & & & & & & & & & & & & & & & \\[-0.2cm]
$\lambda$ [\AA] & Id. & \parbox[c]{0.8cm}{\centering $\chi$ [eV]} & \mc{2}{c}{\loggf} & \mc{2}{c|}{\parbox[c]{0.7cm}{\centering $EW$ [m\AA]}} &
$\lambda$ [\AA] & Id. & \parbox[c]{0.8cm}{\centering $\chi$ [eV]} & \mc{2}{c}{\loggf} & \mc{2}{c|}{\parbox[c]{0.7cm}{\centering $EW$ [m\AA]}} &
$\lambda$ [\AA] & Id. & \parbox[c]{0.8cm}{\centering $\chi$ [eV]} & \mc{2}{c}{\loggf} & \mc{2}{c}{\parbox[c]{0.7cm}{\centering $EW$ [m\AA]}} \\[0.2cm]
\hline
 & & & & & & & & & & & & & & & & & & & & \\[-0.2cm]
5247.06 & \ion{Fe}{i}  & 0.09 & $-$5&.022 &  64&.0 & 6265.14 & \ion{Fe}{i}  & 2.18 & $-$2&.588 &  87&.4 & 5426.24 & \ion{Ti}{i}  & 0.02 & $-$2&.778 &  11&.3 \\
5322.05 & \ion{Fe}{i}  & 2.28 & $-$2&.878 &  62&.6 & 6380.75 & \ion{Fe}{i}  & 4.19 & $-$1&.315 &  53&.6 & 5471.20 & \ion{Ti}{i}  & 1.44 & $-$1&.558 &	7&.7 \\
5501.48 & \ion{Fe}{i}  & 0.96 & $-$3&.105 & 121&.2 & 6498.94 & \ion{Fe}{i}  & 0.96 & $-$4&.608 &  47&.4 & 5490.15 & \ion{Ti}{i}  & 1.46 & $-$0&.994 &  22&.1 \\
5522.45 & \ion{Fe}{i}  & 4.21 & $-$1&.397 &  46&.1 & 6608.03 & \ion{Fe}{i}  & 2.28 & $-$3&.945 &  18&.1 & 5648.57 & \ion{Ti}{i}  & 2.49 & $-$0&.378 &  10&.8 \\
5543.94 & \ion{Fe}{i}  & 4.22 & $-$1&.055 &  64&.7 & 6627.55 & \ion{Fe}{i}  & 4.55 & $-$1&.476 &  28&.5 & 5679.94 & \ion{Ti}{i}  & 2.47 & $-$0&.690 &	5&.9 \\
5546.51 & \ion{Fe}{i}  & 4.37 & $-$1&.098 &  54&.6 & 6703.57 & \ion{Fe}{i}  & 2.76 & $-$3&.010 &  37&.8 & 5739.46 & \ion{Ti}{i}  & 2.25 & $-$0&.802 &	7&.3 \\
5560.22 & \ion{Fe}{i}  & 4.43 & $-$1&.080 &  52&.8 & 6726.67 & \ion{Fe}{i}  & 4.61 & $-$1&.060 &  47&.4 & 5866.45 & \ion{Ti}{i}  & 1.07 & $-$0&.812 &  50&.2 \\
5587.58 & \ion{Fe}{i}  & 4.14 & $-$1&.738 &  32&.0 & 6733.16 & \ion{Fe}{i}  & 4.64 & $-$1&.434 &  26&.9 & 6064.63 & \ion{Ti}{i}  & 1.05 & $-$1&.888 &	9&.3 \\
5618.64 & \ion{Fe}{i}  & 4.21 & $-$1&.310 &  50&.9 & 6750.16 & \ion{Fe}{i}  & 2.42 & $-$2&.638 &  74&.7 & 6091.18 & \ion{Ti}{i}  & 2.27 & $-$0&.422 &  15&.8 \\
5619.60 & \ion{Fe}{i}  & 4.39 & $-$1&.422 &  36&.2 & 6752.71 & \ion{Fe}{i}  & 4.64 & $-$1&.206 &  37&.8 & 6126.22 & \ion{Ti}{i}  & 1.07 & $-$1&.352 &  24&.7 \\
5633.95 & \ion{Fe}{i}  & 4.99 & $-$0&.379 &  69&.0 & 5234.63 & \ion{Fe}{ii} & 3.22 & $-$2&.196 &  88&.1 & 6258.10 & \ion{Ti}{i}  & 1.44 & $-$0&.440 &  52&.0 \\
5635.83 & \ion{Fe}{i}  & 4.26 & $-$1&.598 &  33&.6 & 5325.56 & \ion{Fe}{ii} & 3.22 & $-$3&.237 &  40&.3 & 4568.35 & \ion{Ti}{ii} & 1.22 & $-$2&.796 &  31&.4 \\
5638.27 & \ion{Fe}{i}  & 4.22 & $-$0&.828 &  79&.2 & 5414.07 & \ion{Fe}{ii} & 3.22 & $-$3&.574 &  26&.8 & 4583.42 & \ion{Ti}{ii} & 1.16 & $-$2&.934 &  28&.1 \\
5641.44 & \ion{Fe}{i}  & 4.26 & $-$0&.997 &  66&.6 & 5425.25 & \ion{Fe}{ii} & 3.20 & $-$3&.235 &  41&.5 & 4657.21 & \ion{Ti}{ii} & 1.24 & $-$2&.312 &  52&.9 \\
5649.99 & \ion{Fe}{i}  & 5.10 & $-$0&.818 &  35&.0 & 5991.38 & \ion{Fe}{ii} & 3.15 & $-$3&.568 &  30&.2 & 4798.54 & \ion{Ti}{ii} & 1.08 & $-$2&.664 &  44&.4 \\
5651.47 & \ion{Fe}{i}  & 4.47 & $-$1&.733 &  19&.6 & 6084.11 & \ion{Fe}{ii} & 3.20 & $-$3&.770 &  21&.1 & 5211.54 & \ion{Ti}{ii} & 2.59 & $-$1&.484 &  33&.3 \\
5652.32 & \ion{Fe}{i}  & 4.26 & $-$1&.700 &  28&.8 & 6149.25 & \ion{Fe}{ii} & 3.89 & $-$2&.719 &  36&.5 & 5336.78 & \ion{Ti}{ii} & 1.58 & $-$1&.638 &  72&.0 \\
5653.87 & \ion{Fe}{i}  & 4.39 & $-$1&.388 &  38&.0 & 6247.56 & \ion{Fe}{ii} & 3.89 & $-$2&.290 &  55&.4 & 5381.02 & \ion{Ti}{ii} & 1.57 & $-$1&.866 &  61&.8 \\
5661.35 & \ion{Fe}{i}  & 4.28 & $-$1&.813 &  23&.3 & 6369.46 & \ion{Fe}{ii} & 2.89 & $-$4&.150 &  18&.6 & 5418.76 & \ion{Ti}{ii} & 1.58 & $-$2&.106 &  50&.0 \\
5662.52 & \ion{Fe}{i}  & 4.18 & $-$0&.628 &  95&.5 & 6416.93 & \ion{Fe}{ii} & 3.89 & $-$2&.618 &  41&.1 & 5657.44 & \ion{V}{i}   & 1.06 &     &--   &	5&.8 \\
5667.52 & \ion{Fe}{i}  & 4.18 & $-$1&.299 &  53&.3 & 6432.69 & \ion{Fe}{ii} & 2.89 & $-$3&.545 &  42&.6 & 5668.36 & \ion{V}{i}   & 1.08 &     &--   &	5&.3 \\
5679.03 & \ion{Fe}{i}  & 4.65 & $-$0&.774 &  60&.3 & 6456.39 & \ion{Fe}{ii} & 3.90 & $-$2&.133 &  62&.9 & 5670.85 & \ion{V}{i}   & 1.08 &     &--   &  19&.0 \\
5701.55 & \ion{Fe}{i}  & 2.56 & $-$2&.218 &  84&.5 & 6154.23 & \ion{Na}{i}  & 2.10 & $-$1&.619 &  38&.0 & 5727.66 & \ion{V}{i}   & 1.05 &     &--   &	8&.0 \\
5731.77 & \ion{Fe}{i}  & 4.26 & $-$1&.150 &  57&.6 & 6160.75 & \ion{Na}{i}  & 2.10 & $-$1&.419 &  52&.5 & 6090.22 & \ion{V}{i}   & 1.08 &     &--   &  35&.2 \\
5741.85 & \ion{Fe}{i}  & 4.26 & $-$1&.624 &  32&.5 & 4571.10 & \ion{Mg}{i}  & 0.00 &	 &--   & 111&.6 & 6135.37 & \ion{V}{i}   & 1.05 &     &--   &  11&.5 \\
5752.04 & \ion{Fe}{i}  & 4.55 & $-$0&.890 &  58&.3 & 4730.04 & \ion{Mg}{i}  & 4.34 &	 &--   &  67&.0 & 6150.15 & \ion{V}{i}   & 0.30 &     &--   &  11&.5 \\
5775.08 & \ion{Fe}{i}  & 4.22 & $-$1&.139 &  60&.4 & 5711.10 & \ion{Mg}{i}  & 4.34 &	 &--   & 109&.0 & 6199.19 & \ion{V}{i}   & 0.29 &     &--   &  13&.6 \\
5793.92 & \ion{Fe}{i}  & 4.22 & $-$1&.620 &  34&.7 & 5785.29 & \ion{Mg}{i}  & 5.11 & $-$1&.890 &  51&.0 & 6216.36 & \ion{V}{i}   & 0.28 &     &--   &  36&.4 \\
5806.73 & \ion{Fe}{i}  & 4.61 & $-$0&.895 &  54&.8 & 5684.48 & \ion{Si}{i}  & 4.95 & $-$1&.638 &  62&.8 & 6274.66 & \ion{V}{i}   & 0.27 &     &--   &	9&.8 \\
5809.22 & \ion{Fe}{i}  & 3.88 & $-$1&.617 &  51&.2 & 5690.43 & \ion{Si}{i}  & 4.93 & $-$1&.800 &  52&.9 & 6285.17 & \ion{V}{i}   & 0.28 &     &--   &	9&.4 \\
5814.81 & \ion{Fe}{i}  & 4.28 & $-$1&.835 &  22&.6 & 5701.11 & \ion{Si}{i}  & 4.93 & $-$1&.994 &  40&.4 & 4626.54 & \ion{Mn}{i}  & 4.71 &     &--   &  29&.5 \\
5852.22 & \ion{Fe}{i}  & 4.55 & $-$1&.183 &  41&.5 & 5793.08 & \ion{Si}{i}  & 4.93 & $-$1&.918 &  45&.2 & 4739.11 & \ion{Mn}{i}  & 2.94 &     &--   &  65&.0 \\
5855.08 & \ion{Fe}{i}  & 4.61 & $-$1&.519 &  23&.0 & 6145.02 & \ion{Si}{i}  & 5.61 & $-$1&.420 &  40&.2 & 5004.89 & \ion{Mn}{i}  & 2.92 &     &--   &  12&.2 \\
5856.09 & \ion{Fe}{i}  & 4.29 & $-$1&.558 &  34&.5 & 5581.98 & \ion{Ca}{i}  & 2.52 & $-$0&.780 &  97&.0 & 5394.67 & \ion{Mn}{i}  & 0.00 &     &--   &  78&.6 \\
5862.36 & \ion{Fe}{i}  & 4.55 & $-$0&.423 &  89&.6 & 5590.13 & \ion{Ca}{i}  & 2.52 & $-$0&.829 &  93&.5 & 5399.48 & \ion{Mn}{i}  & 3.85 &     &--   &  39&.2 \\
5905.68 & \ion{Fe}{i}  & 4.65 & $-$0&.787 &  59&.4 & 5867.57 & \ion{Ca}{i}  & 2.93 & $-$1&.566 &  26&.6 & 5413.68 & \ion{Mn}{i}  & 3.86 &     &--   &  23&.5 \\
5916.26 & \ion{Fe}{i}  & 2.45 & $-$2&.922 &  55&.0 & 6161.29 & \ion{Ca}{i}  & 2.52 & $-$1&.362 &  59&.6 & 5420.35 & \ion{Mn}{i}  & 2.14 &     &--   &  83&.6 \\
5927.79 & \ion{Fe}{i}  & 4.65 & $-$1&.068 &  43&.2 & 6163.75 & \ion{Ca}{i}  & 2.52 & $-$1&.295 &  63&.8 & 5432.55 & \ion{Mn}{i}  & 0.00 &     &--   &  51&.6 \\
5929.68 & \ion{Fe}{i}  & 4.55 & $-$1&.207 &  40&.5 & 6166.44 & \ion{Ca}{i}  & 2.52 & $-$1&.188 &  70&.6 & 5537.77 & \ion{Mn}{i}  & 2.19 &     &--   &  36&.4 \\
5930.19 & \ion{Fe}{i}  & 4.65 & $-$0&.336 &  90&.8 & 6169.04 & \ion{Ca}{i}  & 2.52 & $-$0&.842 &  94&.6 & 6013.50 & \ion{Mn}{i}  & 3.07 &     &--   &  87&.0 \\
5934.66 & \ion{Fe}{i}  & 3.93 & $-$1&.139 &  76&.3 & 6169.56 & \ion{Ca}{i}  & 2.52 & $-$0&.621 & 112&.3 & 6021.80 & \ion{Mn}{i}  & 3.07 &     &--   &  91&.2 \\
5983.69 & \ion{Fe}{i}  & 4.55 & $-$0&.839 &  61&.6 & 6455.60 & \ion{Ca}{i}  & 2.52 & $-$1&.415 &  57&.3 & 5032.72 & \ion{Ni}{i}  & 3.90 & $-$1&.178 &  24&.9 \\
5984.82 & \ion{Fe}{i}  & 4.73 & $-$0&.346 &  85&.1 & 6499.65 & \ion{Ca}{i}  & 2.52 & $-$0&.954 &  87&.5 & 5094.41 & \ion{Ni}{i}  & 3.83 & $-$1&.052 &  33&.3 \\
6024.06 & \ion{Fe}{i}  & 4.55 & $-$0&.142 & 113&.8 & 4518.02 & \ion{Ti}{i}  & 0.83 & $-$0&.394 &  74&.2 & 5220.30 & \ion{Ni}{i}  & 3.74 & $-$1&.246 &  28&.7 \\
6027.06 & \ion{Fe}{i}  & 4.08 & $-$1&.195 &  65&.1 & 4548.77 & \ion{Ti}{i}  & 0.83 & $-$0&.438 &  72&.5 & 5392.33 & \ion{Ni}{i}  & 4.15 & $-$1&.328 &  12&.7 \\
6056.01 & \ion{Fe}{i}  & 4.73 & $-$0&.507 &  73&.5 & 4617.25 & \ion{Ti}{i}  & 1.75 & $ $0&.288 &  66&.3 & 5435.87 & \ion{Ni}{i}  & 1.99 & $-$2&.440 &  52&.3 \\
6065.49 & \ion{Fe}{i}  & 2.61 & $-$1&.687 & 119&.6 & 4758.12 & \ion{Ti}{i}  & 2.25 & $ $0&.312 &  44&.5 & 5452.86 & \ion{Ni}{i}  & 3.84 & $-$1&.482 &  16&.6 \\
6079.01 & \ion{Fe}{i}  & 4.65 & $-$1&.015 &  46&.4 & 4759.27 & \ion{Ti}{i}  & 2.25 & $ $0&.340 &  46&.0 & 5494.88 & \ion{Ni}{i}  & 4.10 & $-$1&.055 &  23&.0 \\
6082.72 & \ion{Fe}{i}  & 2.22 & $-$3&.572 &  34&.7 & 4778.26 & \ion{Ti}{i}  & 2.24 & $-$0&.380 &  15&.6 & 5587.85 & \ion{Ni}{i}  & 1.93 & $-$2&.438 &  55&.8 \\
6094.38 & \ion{Fe}{i}  & 4.65 & $-$1&.553 &  20&.6 & 4926.15 & \ion{Ti}{i}  & 0.82 & $-$2&.240 &   6&.1 & 5625.31 & \ion{Ni}{i}  & 4.09 & $-$0&.714 &  39&.3 \\
6096.67 & \ion{Fe}{i}  & 3.98 & $-$1&.788 &  38&.1 & 5022.87 & \ion{Ti}{i}  & 0.83 & $-$0&.452 &  75&.4 & 5637.13 & \ion{Ni}{i}  & 4.09 & $-$0&.846 &  32&.8 \\
6151.62 & \ion{Fe}{i}  & 2.18 & $-$3&.302 &  50&.1 & 5071.47 & \ion{Ti}{i}  & 1.46 & $-$0&.738 &  31&.6 & 6176.81 & \ion{Ni}{i}  & 4.09 & $-$0&.270 &  65&.3 \\
6157.73 & \ion{Fe}{i}  & 4.08 & $-$1&.244 &  63&.2 & 5113.45 & \ion{Ti}{i}  & 1.44 & $-$0&.846 &  27&.8 & 6177.24 & \ion{Ni}{i}  & 1.83 & $-$3&.497 &  15&.8 \\
6165.36 & \ion{Fe}{i}  & 4.14 & $-$1&.488 &  46&.1 & 5145.46 & \ion{Ti}{i}  & 1.46 & $-$0&.626 &  37&.3 & 6186.71 & \ion{Ni}{i}  & 4.10 & $-$0&.886 &  31&.5 \\
6180.21 & \ion{Fe}{i}  & 2.73 & $-$2&.572 &  60&.1 & 5147.48 & \ion{Ti}{i}  & 0.00 & $-$2&.154 &  33&.5 & 6378.26 & \ion{Ni}{i}  & 4.15 & $-$0&.812 &  33&.2 \\
6188.00 & \ion{Fe}{i}  & 3.94 & $-$1&.629 &  48&.5 & 5152.19 & \ion{Ti}{i}  & 0.02 & $-$2&.088 &  35&.7 & 5782.14 & \ion{Cu}{i}  & 1.64 &     &--   &  80&.5 \\
6200.32 & \ion{Fe}{i}  & 2.61 & $-$2&.405 &  74&.7 & 5192.97 & \ion{Ti}{i}  & 0.02 & $-$1&.012 &  87&.9 & 5853.69 & \ion{Ba}{ii} & 0.60 & $-$0&.888 &  63&.3 \\
6226.74 & \ion{Fe}{i}  & 3.88 & $-$2&.064 &  29&.7 & 5211.21 & \ion{Ti}{i}  & 0.84 & $-$2&.120 &   7&.9 & 6141.73 & \ion{Ba}{ii} & 0.70 & $ $0&.043 & 114&.3 \\
6229.24 & \ion{Fe}{i}  & 2.84 & $-$2&.860 &  39&.8 & 5295.78 & \ion{Ti}{i}  & 1.07 & $-$1&.670 &  12&.5 & 	  &		 &	&     &     &	 &   \\
6240.65 & \ion{Fe}{i}  & 2.22 & $-$3&.298 &  48&.7 & 5219.70 & \ion{Ti}{i}  & 0.02 & $-$2&.250 &  28&.6 & 	  &		 &	&     &     &	 &   \\
\hline
\end{tabular}
\tablefoot{Lines with missing $gf$ values represent the elements for which the hyperfine structure was taken into account (Mg, V, Mn, and Cu) and the detailed line splitting is shown in Table~\ref{hfs}.}
\end{table*}

\begin{table*}
\centering
\caption[]{The $gf$ values for lines with hyperfine structure computed based            on the degraded spectrum of the Solar Flux Atlas.}
\label{hfs}
\begin{tabular}{cr | cr | cr}
\hline\hline
 & & & & & \\[-0.2cm]
$\lambda$ [\AA] & \loggf\             & $\lambda$ [\AA] & \loggf\             & $\lambda$ [\AA] & \loggf\            \\[0.1cm]
\hline
 & & & \\[-0.2cm]
\mc{2}{c|}{\bf \ion{Mg}{i} : 4571.10} & \mc{2}{c|}{\bf \ion{V}{i} : 6274.66}  & \mc{2}{c}{\bf \ion{Mn}{i} : 5420.35} \\
4571.078 & $-$6.554		      & 6274.640 & $-$2.150                   & 5420.277 & $-$2.311                  \\
4571.087 & $-$6.594		      & 6274.658 & $-$2.150                   & 5420.301 & $-$2.233                  \\
4571.096 & $-$5.694		      & 6274.676 & $-$2.150                   & 5420.334 & $-$3.093                  \\
\mc{2}{c|}{\bf \ion{Mg}{i} : 4730.04} & \mc{2}{c|}{\bf \ion{V}{i} : 6285.17}  & 5420.376 & $-$1.986                  \\
4730.031 & $-$3.254                   & 6285.147 & $-$2.160		      & 5420.429 & $-$1.898                  \\
4730.038 & $-$3.294                   & 6285.165 & $-$2.160		      & \mc{2}{c}{\bf \ion{Mn}{i} : 5432.55} \\
4730.046 & $-$2.394                   & 6285.183 & $-$2.160		      & 5432.512 & $-$4.348                  \\
\mc{2}{c|}{\bf \ion{Mg}{i} : 5711.10} & \mc{2}{c|}{\bf \ion{Mn}{i} : 4626.54} & 5432.540 & $-$4.434                  \\
5711.074 & $-$2.716                   & 4626.464 & $-$0.778		      & 5432.565 & $-$4.544                  \\
5711.083 & $-$2.756                   & 4626.504 &    0.017		      & 5432.584 & $-$4.689                  \\
5711.091 & $-$1.856                   & 4626.530 & $-$0.231		      & 5432.598 & $-$4.783                  \\
\mc{2}{c|}{\bf \ion{V}{i} : 5657.44}  & 4626.565 &    0.133		      & \mc{2}{c}{\bf \ion{Mn}{i} : 5537.77} \\
5657.418 & $-$1.569                   & 4626.573 & $-$0.327		      & 5537.691 & $-$2.802                  \\
5657.436 & $-$1.569                   & \mc{2}{c|}{\bf \ion{Mn}{i} : 4739.11} & 5537.710 & $-$2.689                  \\
5657.454 & $-$1.569                   & 4739.099 & $-$1.189                   & 5537.738 & $-$2.653                  \\
\mc{2}{c|}{\bf \ion{V}{i} : 5668.36}  & 4739.113 & $-$1.330                   & 5537.764 & $-$2.689                  \\
5668.344 & $-$1.590                   & 4739.126 & $-$1.485                   & 5537.802 & $-$2.325                  \\
5668.362 & $-$1.590                   & 4739.145 & $-$1.042                   & \mc{2}{c}{\bf \ion{Mn}{i} : 6013.50} \\
5668.380 & $-$1.590                   & 4739.167 & $-$2.392                   & 6013.474 & $-$0.704		     \\
\mc{2}{c|}{\bf \ion{V}{i} : 5670.85}  & \mc{2}{c}{\bf \ion{Mn}{i} : 5004.89}  & 6013.486 & $-$0.914		     \\
5670.833 & $-$0.970		      & 5004.878 & $-$2.249		      & 6013.501 & $-$1.045		     \\
5670.851 & $-$0.970		      & 5004.892 & $-$2.390		      & 6013.519 & $-$0.724		     \\
5670.869 & $-$0.970		      & 5004.905 & $-$2.545		      & 6013.537 & $-$1.302		     \\
\mc{2}{c|}{\bf \ion{V}{i} : 5727.66}  & 5004.924 & $-$2.102		      & \mc{2}{c}{\bf \ion{Mn}{i} : 6021.80} \\
5727.643 & $-$1.434		      & 5004.946 & $-$3.452		      & 6021.764 & $-$1.386		     \\
5727.661 & $-$1.434		      & \mc{2}{c}{\bf \ion{Mn}{i} : 5394.67}  & 6021.780 & $-$1.237		     \\
5727.679 & $-$1.434		      & 5394.617 & $-$4.028                   & 6021.797 & $-$0.406		     \\
\mc{2}{c|}{\bf \ion{V}{i} : 6090.22}  & 5394.645 & $-$4.114                   & 6021.806 & $-$0.623		     \\
6090.234 & $-$0.629                   & 5394.670 & $-$4.224                   & 6021.814 & $-$0.488		     \\
6090.216 & $-$0.629                   & 5394.689 & $-$4.369                   & \mc{2}{c}{\bf \ion{Cu}{i} : 5782.14} \\
6090.198 & $-$0.629                   & 5394.703 & $-$4.463                   & 5782.032 & $-$3.554                  \\
\mc{2}{c|}{\bf \ion{V}{i} : 6135.37}  & \mc{2}{c}{\bf \ion{Mn}{i} : 5399.48}  & 5782.042 & $-$3.857                  \\
6135.352 & $-$1.280                   & 5399.435 & $-$0.904                   & 5782.054 & $-$3.156                  \\
6135.370 & $-$1.280                   & 5399.446 & $-$1.094                   & 5782.064 & $-$3.207                  \\
6135.388 & $-$1.280                   & 5399.479 & $-$0.995                   & 5782.073 & $-$3.511                  \\
\mc{2}{c|}{\bf \ion{V}{i} : 6150.15}  & 5399.502 & $-$0.616                   & 5782.084 & $-$2.810                  \\
6150.136 & $-$2.030                   & 5399.536 & $-$1.245                   & 5782.086 & $-$3.156                  \\
6150.154 & $-$2.030                   & \mc{2}{c}{\bf \ion{Mn}{i} : 5413.68}  & 5782.098 & $-$3.156                  \\
6150.172 & $-$2.030                   & 5413.613 & $-$1.826                   & 5782.113 & $-$2.810                  \\
\mc{2}{c|}{\bf \ion{V}{i} : 6199.19}  & 5413.653 & $-$1.031                   & 5782.124 & $-$2.810                  \\
6199.168 & $-$1.960		      & 5413.679 & $-$1.279                   & 5782.153 & $-$2.709                  \\
6199.186 & $-$1.960		      & 5413.714 & $-$0.915                   & 5782.173 & $-$2.363                  \\
6199.204 & $-$1.960		      & 5413.722 & $-$1.375                   &          &                           \\
\mc{2}{c|}{\bf \ion{V}{i} : 6216.36}  &          &                            &          &                           \\
6216.340 & $-$1.416                   &          &                            &          &                           \\
6216.358 & $-$1.416                   &          &                            &          &                           \\
6216.376 & $-$1.416                   &          &                            &          &                           \\
\hline
\end{tabular}
\end{table*}

\begin{table*}[t!]
\centering
\caption[]{Excerpt from the list of 140 dwarf stars with the photospheric            parameters and [Ti/Fe] abundance ratios.}
\label{spec_pars}
\begin{tabular}{lcc r@{ }l cc r@{ }l r@{ }l}
\noalign{\smallskip}\hline\hline\noalign{\smallskip}
Star &
\parbox[c]{1.1cm}{\centering Spectral type} &
\parbox[c]{1.1cm}{\centering \Vbroad\ [\kms]} &
\multicolumn{2}{c}{\parbox[c]{1.0cm}{\centering \teff\ $ \pm\ \sigma$ [K]}} &
\parbox[c]{1.2cm}{\centering \logg\ $\pm\ \sigma$} &
\parbox[c]{1.1cm}{\centering $\xi$ $\pm\ \sigma$ [\kms]} &
\multicolumn{2}{c}{[Fe/H] $\pm\ \sigma$} &
\multicolumn{2}{c}{[Ti/Fe] $\pm\ \sigma$} \\
\noalign{\smallskip}\hline\noalign{\smallskip}
BD+290503  & G5      & 0.00 & 5180 & $\pm$ 50  & 4.26 $\pm$ 0.20 & 0.86 $\pm$ 0.31 &	0.11 & $\pm$ 0.07 & $-$0.07 & $\pm$ 0.10 \\
HD\,10145  & G5\,V   & 0.00 & 5593 & $\pm$ 43  & 4.25 $\pm$ 0.19 & 0.78 $\pm$ 0.14 &	0.01 & $\pm$ 0.07 &    0.03 & $\pm$ 0.07 \\
HD\,10307  & G1.5\,V & 0.00 & 5913 & $\pm$ 49  & 4.29 $\pm$ 0.22 & 1.06 $\pm$ 0.09 &	0.09 & $\pm$ 0.06 & $-$0.02 & $\pm$ 0.07 \\
HD\,10476  & K1\,V   & 0.00 & 5141 & $\pm$ 31  & 4.29 $\pm$ 0.18 & 0.40 $\pm$ 0.25 & $-$0.05 & $\pm$ 0.05 &    0.02 & $\pm$ 0.08 \\
HD\,106116 & G4\,V   & 0.00 & 5620 & $\pm$ 36  & 4.22 $\pm$ 0.17 & 0.84 $\pm$ 0.11 &	0.16 & $\pm$ 0.05 & $-$0.01 & $\pm$ 0.09 \\
HD\,106516 & F9\,V   & 0.00 & 6401 & $\pm$ 158 & 4.85 $\pm$ 0.39 & 1.59 $\pm$ 0.66 & $-$0.47 & $\pm$ 0.12 &    0.29 & $\pm$ 0.17 \\
HD\,10780  & K0\,V   & 0.00 & 5339 & $\pm$ 26  & 4.40 $\pm$ 0.15 & 0.71 $\pm$ 0.13 &	0.00 & $\pm$ 0.04 &    0.02 & $\pm$ 0.06 \\
HD\,108954 & F9\,V   & 0.00 & 6194 & $\pm$ 60  & 4.60 $\pm$ 0.23 & 1.15 $\pm$ 0.13 &	0.08 & $\pm$ 0.08 &    0.03 & $\pm$ 0.09 \\
HD\,109358 & G0\,V   & 0.00 & 5875 & $\pm$ 30  & 4.42 $\pm$ 0.14 & 1.21 $\pm$ 0.08 & $-$0.22 & $\pm$ 0.04 &    0.03 & $\pm$ 0.06 \\
HD\,11007  & F8\,V   & 0.00 & 6032 & $\pm$ 44  & 4.16 $\pm$ 0.26 & 1.39 $\pm$ 0.09 & $-$0.14 & $\pm$ 0.06 &    0.03 & $\pm$ 0.09 \\
...        & ...     & ...  &	   & ...       & ...		 & ...  	   &	     & ...	  &	    & ...	 \\  
\hline
\end{tabular}
\tablefoot{The broadening velocity \Vbroad\ is also shown. The whole tables for this and other elements, also for subgiants and giants, are available in electronic form at the CDS.}
\end{table*}

\addtocounter{table}{2}
\begin{table*}[t!]
\centering
\caption[]{The same as Table~\ref{spec_pars} but showing photometric and            evolutionary parameters together with the population membership.}
\label{evol_pars}
\begin{tabular}{lcc r@{}l r@{}l r@{ }l c r@{}l ccc}
\noalign{\smallskip}\hline\hline\noalign{\smallskip}
Star & \bv & $M_{\rm V}$ &
\multicolumn{2}{c}{\parbox[c]{0.6cm}{\centering dist. [pc]}} &
\multicolumn{2}{c}{\parbox[c]{0.5cm}{\centering $BC$}} &
\multicolumn{2}{c}{\parbox[c]{0.7cm}{\centering $\log{L}$}} &
\parbox[c]{0.7cm}{\centering mass [\Msun]} &
\multicolumn{2}{c}{\parbox[c]{0.7cm}{\centering age [Gyr]}} &
\parbox[c]{1.4cm}{\centering Population group} &
Remarks & Ref. \\
\noalign{\smallskip}\hline\noalign{\smallskip}
HD\,10145  & 0.691 & 4.84 & 37&.3 & $-$0&.106 &    0.005 & $\pm$ 0.026 & 0.94 $\pm$ 0.03 & 10&${.1^{+2.5}}_{-1.8}$ & thin	&              & \\
HD\,10476  & 0.836 & 5.82 &  7&.5 & $-$0&.239 & $-$0.331 & $\pm$ 0.006 & 0.78 $\pm$ 0.04 & 17&${.1^{+8.2}}_{-7.7}$ & thin	& $a$          & \\
HD\,106116 & 0.701 & 4.79 & 34&.9 & $-$0&.100 &    0.025 & $\pm$ 0.015 & 1.00 $\pm$ 0.03 &  7&${.1^{+2.5}}_{-1.4}$ & thin/thick &              & \\
HD\,106516 & 0.470 & 4.36 & 22&.4 &    0&.011 &    0.150 & $\pm$ 0.016 & 1.05 $\pm$ 0.05 &  0&${.2^{+1.3}}_{-0.2}$ & thick	&              & \\
HD\,108954 & 0.568 & 4.53 & 21&.8 & $-$0&.010 &    0.092 & $\pm$ 0.008 & --              &   &--                   & thin	& $b$          & \\
HD\,128311 & 0.973 & 6.38 & 16&.6 & $-$0&.126 & $-$0.603 & $\pm$ 0.015 & --              &  0&.50                  & thin	& $c$, $d$ (U) & 1,2 \\
HD\,124292 & 0.733 & 6.18 & 22&.1 & $-$0&.143 & $-$0.516 & $\pm$ 0.011 & --              &   &--                   & thin	& $e$          & 3 \\
HD\,206860 & 0.587 & 4.74 & 17&.9 & $-$0&.020 &    0.017 & $\pm$ 0.008 & --              &  0&.10                  & thin	& $c$, $d$ (P) & 1,2 \\
HD\,25825  & 0.593 & 4.50 & 45&.9 & $-$0&.021 &    0.108 & $\pm$ 0.052 & --              &  0&.65                  & thin	& $d$ (H)      & 4 \\
HD\,42807  & 0.663 & 5.17 & 18&.0 & $-$0&.070 & $-$0.140 & $\pm$ 0.008 & --              &  2&.70                  & thin	& $d$ (W)      & 5 \\
...	   & ...   & ...  & ..&.  &     & ... &          & ...         & ...		 &   & ...                 & ...	&              & \\ 
\hline
\end{tabular}
\tablefoot{
\tablefoottext{a}{Not considered in the abundance analysis because the age uncertainties are larger than 4~Gyr (see Sect.~\ref{mass_age});}
\tablefoottext{b}{Mass and age not derived because the star is located outside the evolutionary tracks and isochrones used;}
\tablefoottext{c}{Star hosting at least one giant planet;}
\tablefoottext{d}{Member of a stellar moving group (U: Ursa Major; P: Pleiades; H: Hyades; W: Wolf 630);}
\tablefoottext{e}{Subdwarf candidate.}
}
\tablebib{
\tablefoottext{1}{http://exoplanet.eu};
\tablefoottext{2}{\citet{Montesetal2001}};
\tablefoottext{3}{\citet{SandageFouts1986}};
\tablefoottext{4}{\citet{Eggen1950}};
\tablefoottext{5}{\citet{BubarKing2010}}
}
\end{table*}

%
%
%
%

\end{document}